\documentclass[journal]{IEEEtran}
\usepackage{amssymb}
\usepackage{graphicx}
\usepackage[cmex10]{amsmath}
\usepackage{subfigure}
\usepackage{times}
\usepackage{comme nt}
\usepackage{multirow}
\usepackage{ifpdf}
\usepackage{amsfonts}
\usepackage{txfonts}
\usepackage{cite}
\usepackage{listings}
\usepackage{xcolor}
\usepackage{url}

\begin{document}
%
\title{Q-STAR:A Perceptual Video Quality Model Considering Impact of Spatial, Temporal, and Amplitude Resolutions}

%
%


\author{Yen-Fu Ou,
	    Yuanyi Xue,
	    Yao Wang\thanks{Y.-F. Ou, Y. Xue, and Y. Wang are with the Polytechnic Institute of New York
University, Brooklyn, NY 11201 USA (e-mail: you01@students.poly.edu;
yxue01@students.poly.edu; yao@poly.edu).}
}
	



\maketitle

\begin{abstract}
In this paper, we investigate the impact of spatial, temporal and amplitude resolution (STAR) on the perceptual quality of a compressed video. Subjective quality tests were
carried out on a mobile device. Seven source sequences are included in the tests and for each source sequence we have 27 test configurations generated by JSVM encoder (3 QP levels, 3 spatial resolutions, and 3 temporal resolutions), resulting a total of 189 processed video sequences (PVSs). Videos coded at different spatial resolutions are displayed at the full screen size of the mobile platform. Subjective data reveal that the impact of spatial resolution (SR), temporal resolution (TR) and quantization stepsize (QS) can each be captured by a function with a single content-dependent parameter. The joint impact of SR, TR and QS can be accurately modeled by the product of these three functions with only three parameters. We further find that the quality decay rates with SR and QS, respectively are independent of TR, and likewise, the decay rate with TR is independent of SR and QS, respectively. However, there is a significant interaction between the effects of SR and QS. The overall quality model is further validated on five other datasets with very high accuracy. The complete model correlates well with the subjective ratings with a Pearson Correlation Coefficient (PCC) of $0.991$.

\end{abstract}


%

\section{Introduction}\label{sec:intro}
\PARstart{N}{etworked} video applications such as streaming and chat are becoming prevalent, especially over wireless networks and using small mobile devices.
The users in such application are very heterogeneous in their access link bandwidth, processing and display capabilities. The primary parameters of a video bitstream, which control the
bandwidth requirement, include quantization stepsize (QS) (controlling amplitude resolution), frame rate (controlling temporal resolution or TR)
and frame size (controlling spatial resolution or SR). Given the bandwidth limitation and display resolution of a receiver, the encoder, a network transcoder or adaptor has to decide at which spatial, temporal,
and amplitude resolution (STAR) to code, transcode or adapt a video, to achieve
the best perceptual quality. Therefore, it is important to understand the
impact of the STAR on the perceptual quality. On the other hand, studying the joint impact of
all three dimensions on the perceptual quality is a complex and challenging task.

Some prior works e.g.,~\cite{frameVSquantization,Hauske03subjectiveimage,QualityMetricTQuantBitRate, Lee_VideoQualityExperience, cheon, demin, CrossDQualtiyLowBitVideo, Elsevier_08_HWu}, have explored the impact of SR, TR and quantization artifacts (not QS directly) fully or partially on perceived video quality. Among them,~\cite{frameVSquantization} considers the tradeoff between quantization artifacts and frame rate reduction, and the influence of the display environment on this tradeoff. The authors found that high spatial quality is more preferable than high frame rate, but viewers are more sensitive to frame rate reduction for mobile than desktop devices. In~\cite{Hauske03subjectiveimage, QualityMetricTQuantBitRate} authors proposed a video quality metric, which is a wighted sum of two terms, one is PSNR of the interpolated sequences from the original low frame-rate video and another one is frame-rate reduction. However, \cite{Hauske03subjectiveimage} uses constant values and~\cite{QualityMetricTQuantBitRate} uses motion of the sequences as weighting factors. In~\cite{Elsevier_08_HWu} authors proposed to model the quality of a video with varying frame rate as the product of a function of the  frame rate and a function of the QS, which is quite close to our proposed model (when the SR is fixed) conceptually. However they assume the first function is linear with the frame rate, which does not capture the effect of frame rate accurately based on our subjective test results. The second function decays exponentially with the QS, similar to our earlier model for the quantization effect~\cite{yao2}.

Authors in~\cite{demin, cheon, Lee_VideoQualityExperience, CrossDQualtiyLowBitVideo} further consider the influences of spatial resolution on perceived quality. The quality model in~\cite{Lee_VideoQualityExperience} is a function of the bitrate and a so called truncated bitrate ratio of SNR-scalability, while the quality model in~\cite{cheon}, similar to the work in~\cite{Hauske03subjectiveimage}, is a function of PSNR, TR and SR.
Although the quality assessment in~\cite{cheon,demin} include 3 and 6 different spatial resolutions, respectively, these works only involve an SR range from QCIF to CIF. The works in~\cite{Lee_VideoQualityExperience,CrossDQualtiyLowBitVideo} only include two SR's, QCIF and CIF. Furthermore, none of the tests reported in~\cite{demin, cheon, Lee_VideoQualityExperience, CrossDQualtiyLowBitVideo} were carried out on mobile devices. Even though the work in~\cite{Hauske03subjectiveimage} is for mobile devices, authors assumed that model parameters are independent of video content, while on the contrary, authors in~\cite{Elsevier_08_HWu} believe that the model parameters depend on video contents. Nevertheless,~\cite{Elsevier_08_HWu} does not consider how to estimate the model parameters.

In our previous works~\cite{yenfu_csvt, yao2}, we investigated the impact of TR and QS on perceptual video quality, which was evaluated on larger-screen of laptop monitor, and proposed the video quality model considering the effect of TR and QS under a fixed SR (CIF).
In a preliminary study~\cite{yuan_thesis}, we conducted a subjective test to explore the impact of SR and QS, where the subjective tests were done on a mobile platform with small display screen. In this paper, we extend the previous works by considering the interaction of SR, TR, and QS, and propose a complete quality model in terms of SR, TR, and QS. Preliminary results of this study were reported in~\cite{yenfu_ivmsp}. This journal version provides a more comprehensive analysis of the subjective test results, including examining  interaction of the TR, SR, and QS, validating the model with other datasets, and furthermore investigates the quality comparison between scalable videos coded using H.264/SVC and non-scalable videos produced by H.264/AVC.
The remainder of this paper is organized as follows:
Section~\ref{sec:qualitytests} introduces the quality assessment environment, test methodology and data post-processing.
Section~\ref{sec:QSTAR} analyzes the results of subjective tests and present our proposed model. We further study the statistical significance of STAR variables on subjective quality in Sec.~\ref{sec:ANOVA_analysis} and Sec.~\ref{sec:Validation_model} validates the proposed models on other datasets.
We conclude our work in Section~\ref{sec:conclusion}.
\begin{table}[htp]
\centering
\caption{List of Acronyms} \label{tab:acronyms}
\begin{tabular}{|l|p{6 cm}|}
 \hline
TR & Temporal resolution\\
SR & Spatial resolution \\
QP & Quantization parameter\\
QS & Quantization stepsize \\
FR & Frame rate \\
PVS & Processed video sequence\\
MOS & Mean opinion score \\
NQT & Normalized quality v.s. normalized temporal resolution\\
NQS & Normalized quality v.s. normalized spatial resolution \\
NQQ & Normalized quality v.s. inverted normalized quantization stepsize\\
MNQQ & Model for NQQ\\
MNQS & Model for NQS \\
MNQT & Model for NQT \\
PCC & Pearson Correlation Coefficient \\
RMSE & Root mean square error\\
 \hline
\end{tabular}
\end{table}

\section{Testing Platform and Methodology}\label{sec:qualitytests}
\subsection{Testing Platform}
Targeting for wireless mobile applications, we choose TI's Zoom2 mobile development platform (MDP)~\cite{zoom2} as our test platform. This MDP runs on powerful TI OMAP34x processor with a 4.1-inch WVGA (854$\times$480) resolution capacitive multi-touch screen. Google's Android~\cite{android} mobile operating system (OS) version 2.1 (Eclair) is used for our test interface development. Our approach for constructing the interface is using Java and XML code to control the high-level program flow, with the help of Android's SDK library to operate low-level video decoding process.

    \begin{figure}[htp]
    \vspace{-.05in}
    \centering
    \subfigure[{\it Welcome Screen}]{\includegraphics[scale=0.5]{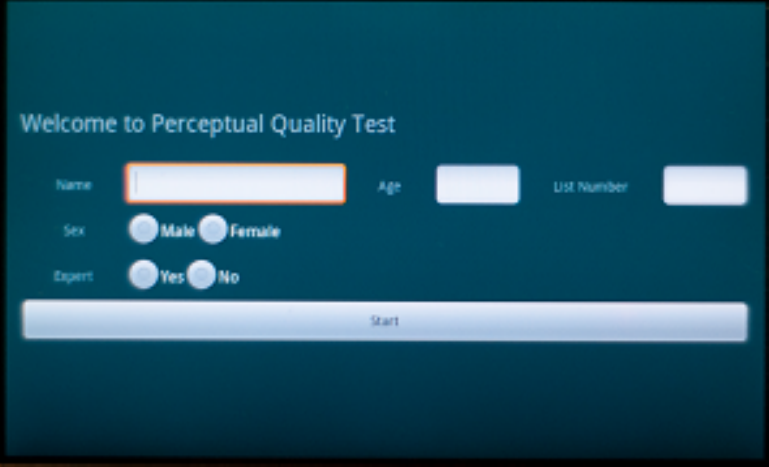}}
    \subfigure[{\it Playback Screen}]{\includegraphics[scale=0.5]{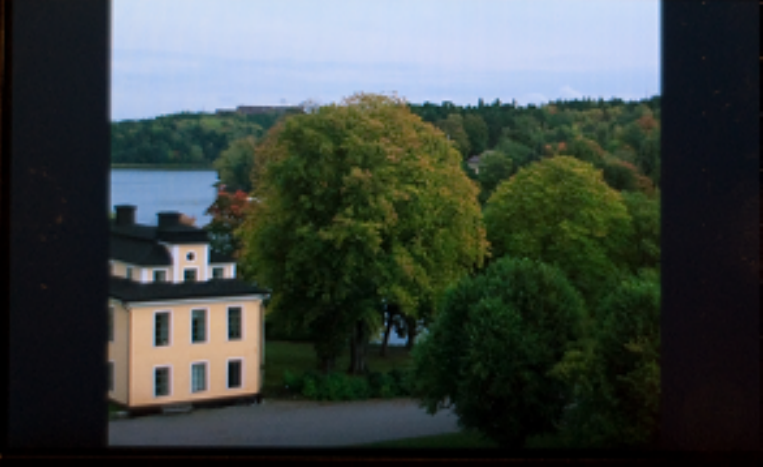}}
    \subfigure[{\it Rating Screen}]{\includegraphics[scale=0.5]{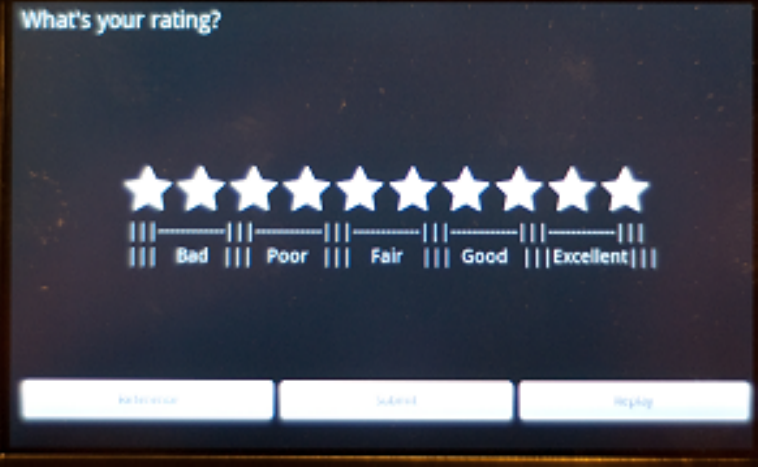}}
    \caption{Screenshots of the subjective rating interface on TI Zoom2 MDP.}
    \label{F:UIdes}
    \vspace{-.05in}
    \end{figure}

\begin{figure*}[htp]
  \centering
  \subfigure[City@4CIF]{\includegraphics[scale=0.007]{city.pdf}}
  \subfigure[Crew@4CIF]{\includegraphics[scale=0.007]{crew.pdf}}
  \subfigure[Harbour@4CIF]{\includegraphics[scale=0.007]{harbour.pdf}}
  \subfigure[Ice@4CIF]{\includegraphics[scale=0.007]{ice.pdf}}
  \subfigure[Soccer@4CIF]{\includegraphics[scale=0.007]{soccer.pdf}}
  \subfigure[FlowerGarden@VGA]{\includegraphics[scale=0.007]{flowerGarden.pdf}}
  \subfigure[Foreman@VGA]{\includegraphics[scale=0.007]{foreman.pdf}}
  \subfigure[Shields@4CIF]{\includegraphics[scale=0.325]{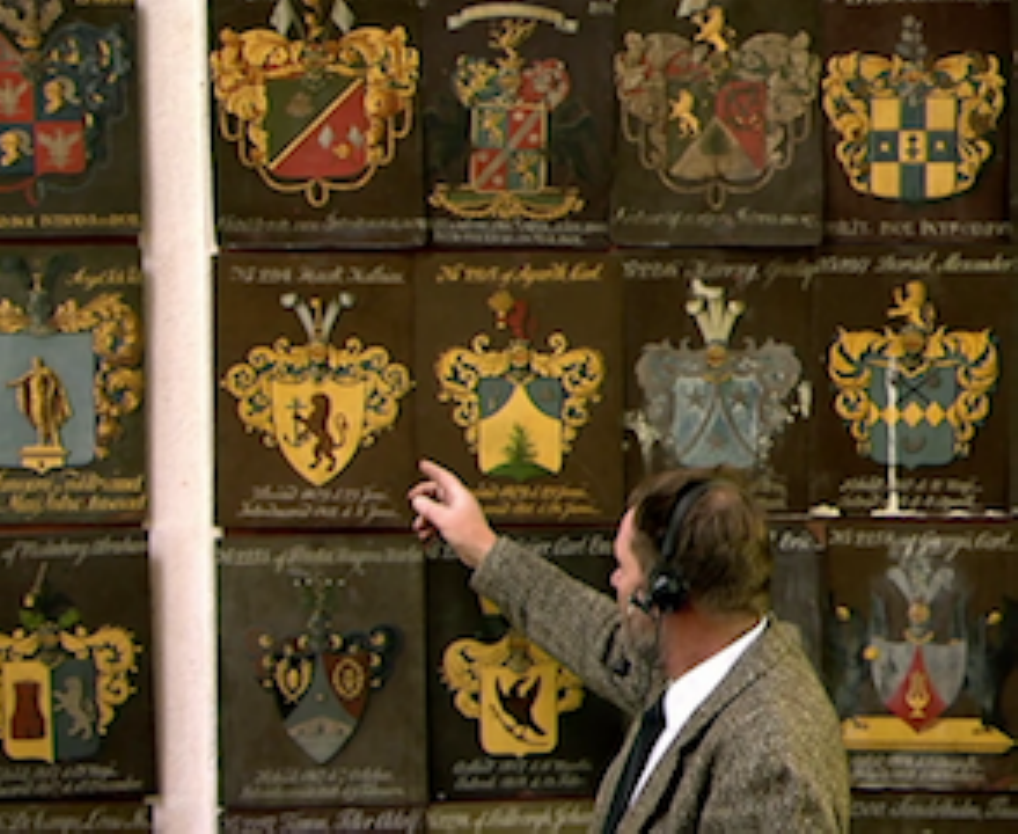}}
  \subfigure[InToTree@4CIF]{\includegraphics[scale=0.325]{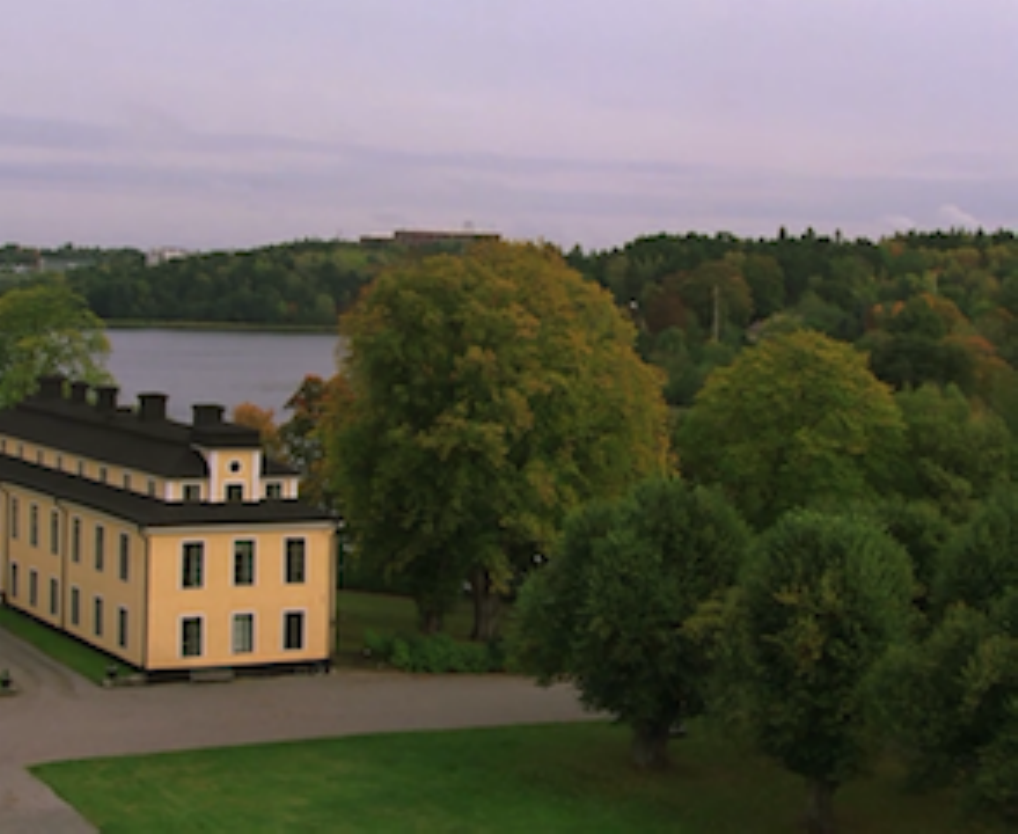}}
  \subfigure[Football@VGA]{\includegraphics[scale=0.32]{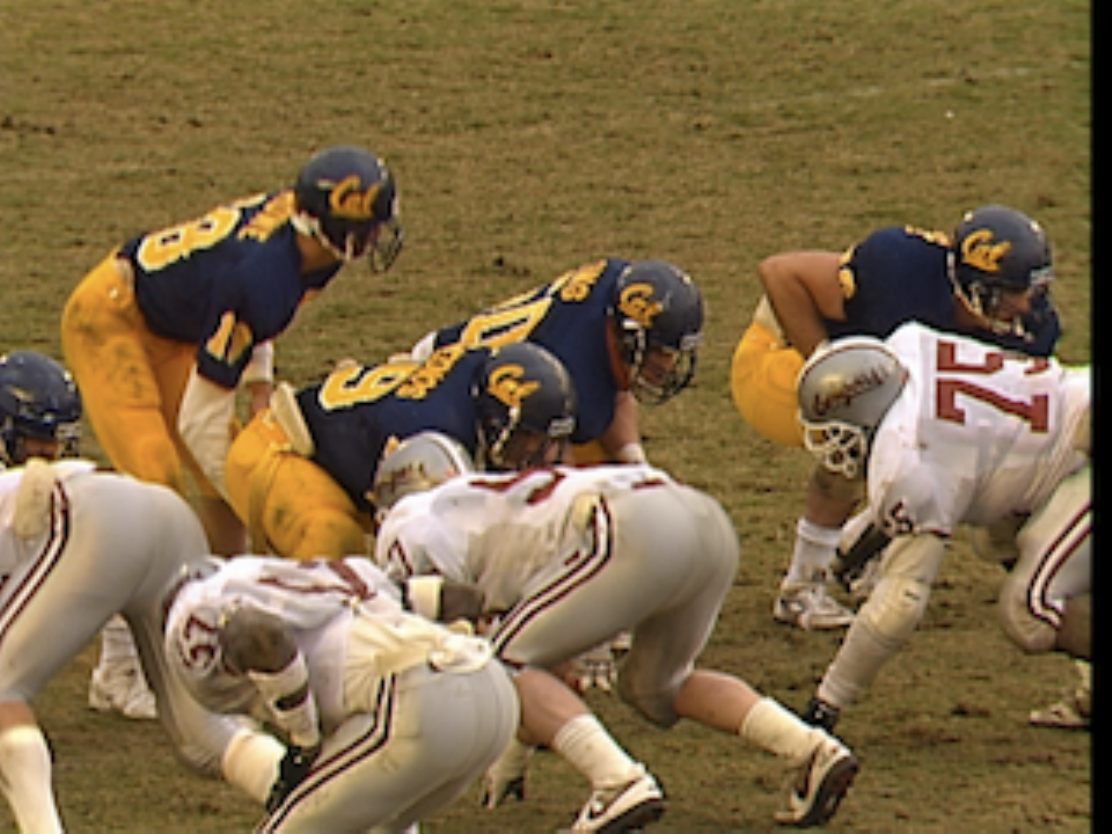}}
  \vspace{-.05in}
  \caption{Test video pool for subjective tests.}
  \label{fig:test_videos}
  \vspace{-.15in}
\end{figure*}

Figure ~\ref{F:UIdes} illustrates subjective rating interface on our Zoom2 MDP. A {\it welcome screen}
is shown to each viewer at the beginning of each test to record his/her basic information (name, age and gender) and then this is followed by a {\it playback screen} on which a random 8-second processed video sequence (PVS) is played. Each viewer will be asked to give a score on a {\it rating screen} after a PVS is played completely. In all tests we allow each subject to replay the current PVS if he/she doesn't feel confident to give a proper judgement, so as to assure more reliable subjective ratings.
We adopted a 10-level rating scale as shown in Fig.~\ref{F:UIdes} (c). We did not put a level below the scale ``1'', which would correspond to a ``totally useless video'', since a viewer can still understand the video scene content even from the video at the lowest STAR in our test video pool. So it is reasonable to interpret the effective rating scale as being 11 levels, as recommended by ITU P.910~\cite{p910}.

\subsection{Test Video Pool} \label{ssec:test_video_pool}
\begin{figure}[htp]
\centering
 \includegraphics[scale=0.7]{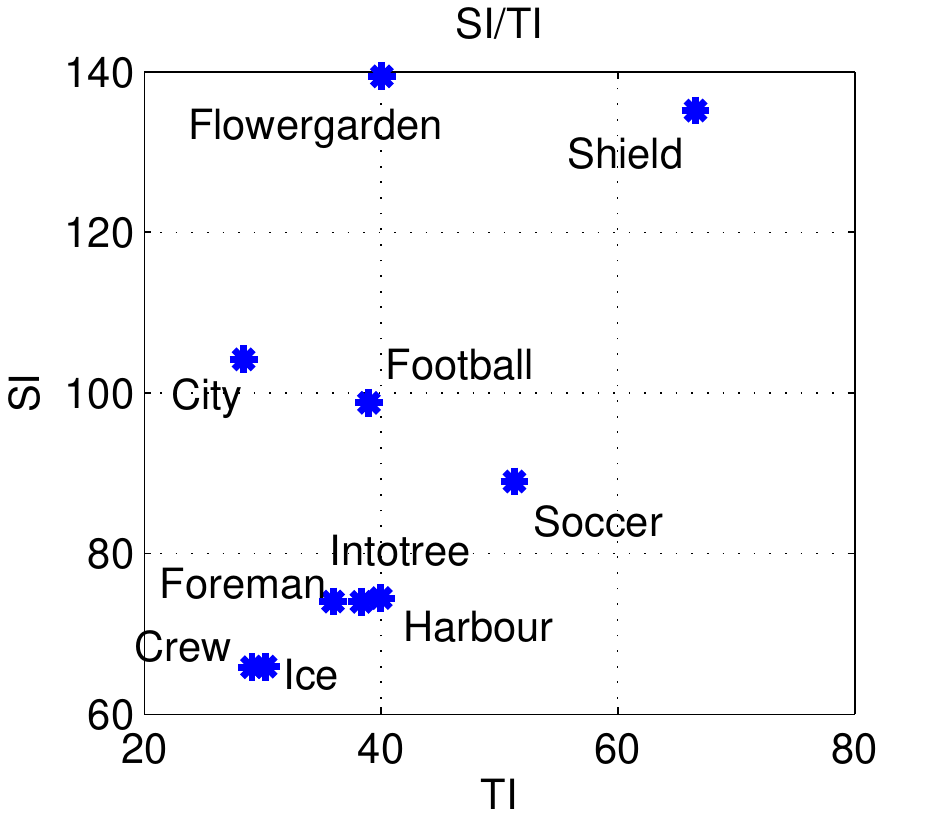}
\caption{The spatial and temporal information indices of the test sequences}\label{fig:SI_TI}
\end{figure}

Ten different videos, i.e., \emph{InToTree}, \emph{Shields}, \emph{Football}, \emph{City}, \emph{Crew}, \emph{Harbour}, \emph{Ice}, \emph{Soccer}, \emph{FlowerGarden} and \emph{Foreman}, two at 720p high-definition (HD), five at 4CIF (704$\times$576) and three at VGA (640$\times$480) resolution, are included in our subjective tests. The first two are cropped from original 720p high-definition (HD) source to match our
Zoom2 MDP display screen size. These videos are selected from the
standard video pool to include various content activities.
We present the snapshot of all source sequences in Fig.~\ref{fig:test_videos} and plot the spatial information (SI) and temporal information (TI) indices~\cite{p910} of all source sequences in Fig.~\ref{fig:SI_TI}. It demonstrates that the test sequence pool covers a wide range of video contents in terms of motion and spatial details. For the testing consistency through all the PVSs, those VGA videos are cropped and interpolated to 4CIF before sending it to the encoder. According to our pretest~\cite{yuan_thesis} performed on the same test platform, it suggests that VGA derived 4CIF versions and original 4CIF versions of the same videos acquire very similar viewer ratings.
Low-resolution (i.e., CIF, QCIF) source videos are obtained by downsampling using
the Sine-waved Sinc function~\cite{downsampling} recommended in the SVC reference software JSVM~\cite{JSVM}. Each source video is encoded by JSVM918~\cite{JSVM} using combined spatial and temporal scalabilities, with 3 spatial layers (4CIF, CIF, QCIF) and 3 temporal layers (30, 15, 7.5Hz). Videos corresponding to different QS's are obtained by coding at different QP's without QP cascading. The GOP size is set to 8 with only the first frame as in the I mode. Hierarchical-B structure is used to provide temporal scalability. The spatial scalability is achieved via the multi-layer coding approach with adaptive
inter-layer prediction. For motion estimation, we use SAD (Sum of Absolute Difference) as cost function for both full-pel and sub-pel. The FastSearch mode is enabled with maximum search range of 16 for full-pel search. The entropy coding method is CAVLC. The other encoding configurations follow the default settings in JSVM.

For display, each PVS under 4CIF resolution is interpolated to 4CIF using the AVC 6-tap half-pel with bilinear quarter-pel interpolation filter~\cite{interpolation}.
The test interface will then automatically resize these 4CIF sequences to a spatial resolution with 480 rows, keeping the aspect ratio of input videos (for 4CIF videos, it is 1.22) by adding the grey-out border on the left and right side. Each PVS is played back in its native frame rate without temporal interpolation.

\subsection{Test Protocol}\label{TP}
Three separate experiments were carried out. Test 1 focuses on the perceptual impact of SR; Test 2 focuses on joint impact of SR and QP; Test 3 focuses on joint effects of STAR. In order to combine subjective scores from these three tests, we include several common sequences between three tests. Common sequences are selected such that they represent a broad quality range in order to facilitate a valid and robust mapping between the tests when combining the datasets. Table~\ref{tab:spatial_res_qp} lists the testing configurations for the three tests. Table~\ref{tab:comm_seq} lists all the common sequences.
Note that for each source video, we tested all combinations of three SR's (QCIF, CIF, 4CIF), three TR's (7.5, 15, 30 Hz) and three QP's (28, 36, 44). Five SR's are examined at  TR=30Hz, QP=22, to allow us to examine the impact of SR in fine granularity when temporal and amplitude resolution are highest. Because our preliminary tests found that videos coded at QP=22 are visually very similar to those at QP=28, QP=22 is not tested at other TR's in Test3.  In deriving the overall quality model, we only use test results for the 27 test conditions (3 QP's, 3 SR's, 3 TR's).


%

\begin{table}[htp]
\vspace{-.in}
\centering
\caption{STAR parameters used in tests}
\label{tab:spatial_res_qp}
\small{
\begin{tabular}{|@{~}c@{~}|c|c|c|}
\hline
  & SR & QP & TR\\
\hline
  \multirow{3}{*}{Test 1} & 176x144 (QCIF), 256x208, & \multirow{3}{*}{22} & \multirow{3}{*}{30} \\
          & 352x288 (CIF), 528x432, & & \\
          &  704x576 (4CIF) & & \\
          \hline
  \multirow{2}{*}{Test 2} & \multirow{2}{*}{QCIF, CIF, 4CIF} &  22, 28, & \multirow{2}{*}{30}\\
    & & 36, 44 & \\
  \hline
  \multirow{2}{*}{Test 3} & \multirow{2}{*}{QCIF, CIF, 4CIF} & 22, 28,  & \multirow{2}{*}{30, 15, 7.5}\\
    & & 36, 44 & \\
  \hline
\end{tabular}}
\vspace{-.in}
\end{table}

\begin{table}[htp]
\vspace{-.0in}
\centering
\caption{Common Sequences}
\label{tab:comm_seq}
\small{
\begin{tabular}{|@{~}c@{~}|@{~}c@{~}|}
\hline
  \multirow{4}{*}{Test 1\&2} & \emph{City}@QP22/4CIF/30Hz, \emph{City}@QP22/QCIF/30Hz,\\
          & \emph{Crew}@QP22/CIF/30Hz, \emph{Harbour}@QP22/QCIF/30Hz,\\
          &  \emph{Ice}@QP22/CIF/30Hz, \emph{Soccer}@QP22/4CIF/30Hz,\\
	    &  \emph{Fg}@QP22/QCIF/30Hz, \emph{Foreman}@QP22/CIF/30Hz.\\
          \hline
  \multirow{2}{*}{Test 2\&3} & For all video contents, QP22/4CIF,  QP28/4CIF\&QCIF,\\
& QP36/CIF, QP44/4CIF\&QCIF, all at 30Hz.\\
  \hline
\end{tabular}}
\vspace{-.15in}
\end{table}

\emph{Single Stimulus}, as recommended by~\cite{p910} is used for all tests. Before the testing session, a training session, which allows viewers to get familiar with the test, is employed. Three source sequences, i.e., \emph{InToTree}, \emph{Shields} and \emph{Football}, are selected as training sequences, and the rest 7 are for testing session.  In Test 2 and 3, we design several subsessions with overlapping sequences, to reduce the viewing time of each subject. Each viewer can participate in one or more subsessions. On average, each viewer spends about 18-20 minutes in one viewing session.

\subsection{Data Processing}\label{DP}

\subsubsection{Data Collection}\label{sssec_DC}
We have around 60 evenly distributed male and female viewers who participated in the tests. Each PVS is rated by 18-20 different viewers. All viewers have normal visual (or after correction) and color perception. About $80\%$ of viewers are non-expert with no related background in video processing. The raw ratings are converted to Z-scores~\cite{AM_zscore} based on the mean and standard deviation of all the scores of each viewer, given by

 \begin{equation}\label{z-score}
    Z_{mij} = \frac{X_{mij} - \mathrm{MEAN}(X_{i})}{\mathrm{STD}(X_{i})}.
 \end{equation}
Here, $X_{mij}$ and $Z_{mij}$ denote the raw rating and Z-score of $m^{th}$ sequence at $j^{th}$ STAR combination, from $i^{th}$ viewer, respectively. $X_{i}$ denotes all ratings from $i^{th}$ viewer. $\mathrm{MEAN}(\cdot)$ and $\mathrm{STD}(\cdot)$ represent the operator for taking the mean and the standard deviation of a given set, respectively.

\subsubsection{Post Screening}\label{sssec_PS}
Two post screening methods are used in concatenation. We first perform BT.500-11 post screening method~\cite{bt500} in Z-score domain to remove all ratings by certain viewers because their ratings are outside the range of the majority of the viewers. On average, one viewer is eliminated for each PVS. We then conduct the second step to the remaining ratings in the raw score domain, using a ratio/averaging method. We make use of the fact that a video coded at a lower SR under the same TR and QP would not have a rating higher than a video coded at a higher SR under the same TR and QP, if the viewer's judgement is consistent. Therefore, we calculate the ratio of ratings by the same viewer for each pair of PVS's with adjacent SR, under the same QP and TR. For each source video and each viewer, we count the number of times that the ratio is greater than a threshold ($=1.1$) for all possible pairs, and we remove all the ratings by a viewer for the same source video if the outlier counter is larger than $2$. For the remaining pairs of ratings by each viewer, if the ratio is larger than $1$, we replace both ratings by their average. We repeat the same procedure for all possible pairs of TR (under the same SR and QP), and for all possible pairs of QP (under the same SR and TR). After this step, approximately 16-18 ratings remain for each PVS.

\subsubsection{Datasets Combining}\label{sssec_Comb}
After the post-processing, we map all the Z-scores from Test 1 and Test 3 to Test 2 using the method recommended in~\cite{Wolf_combinedFunc}. We map all other tests to Test 2 based on the consideration that only Test 2 has a sufficient number of common sequences with both Test 1 and Test 3.

%

\begin{figure*}[!ht]
\vspace{-.in}
\centering
\includegraphics[scale=0.5]{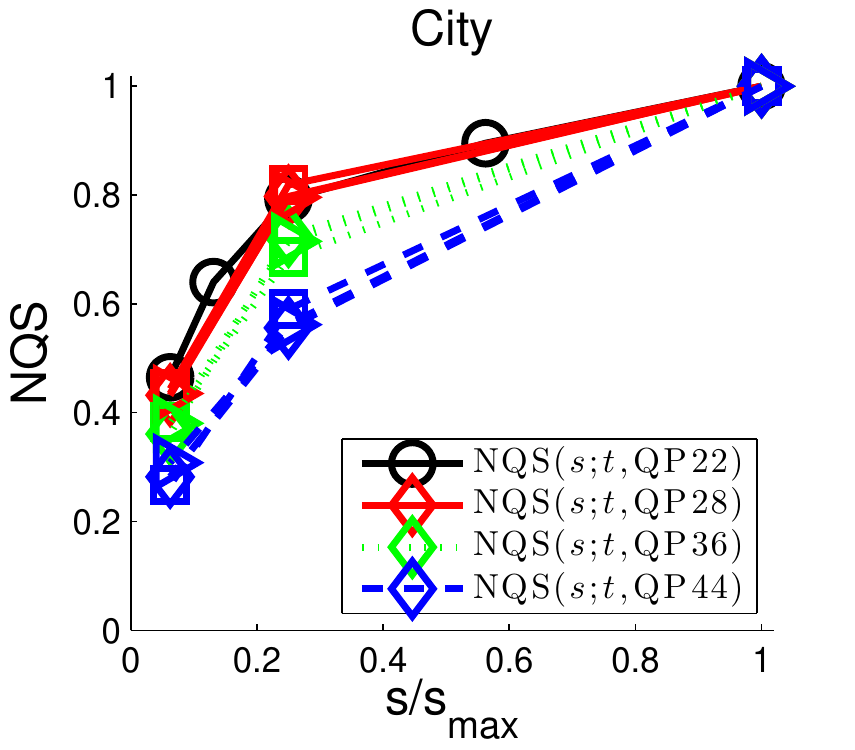}
\includegraphics[scale=0.5]{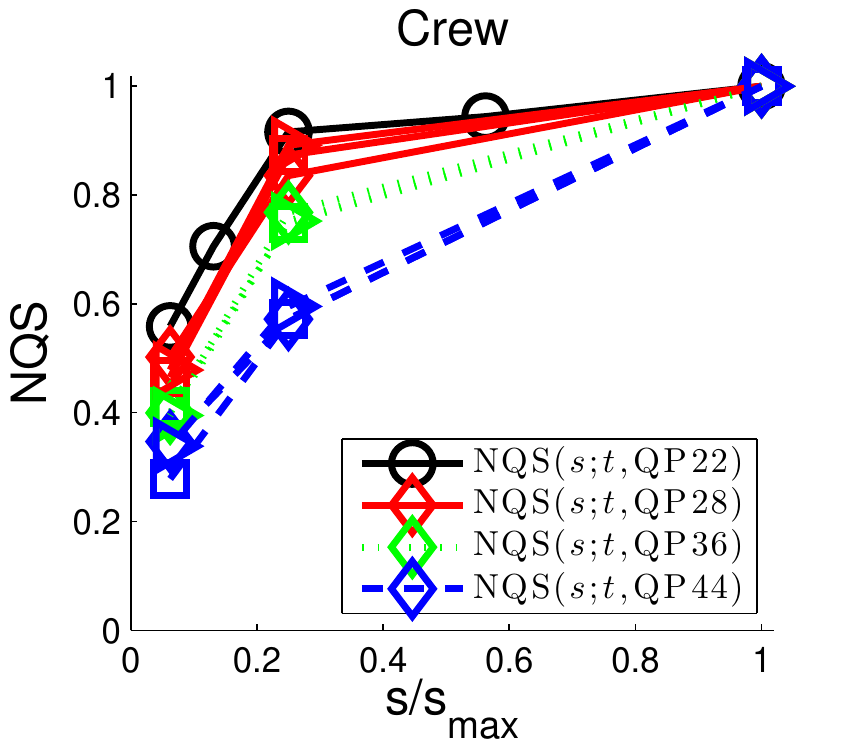}
\includegraphics[scale=0.5]{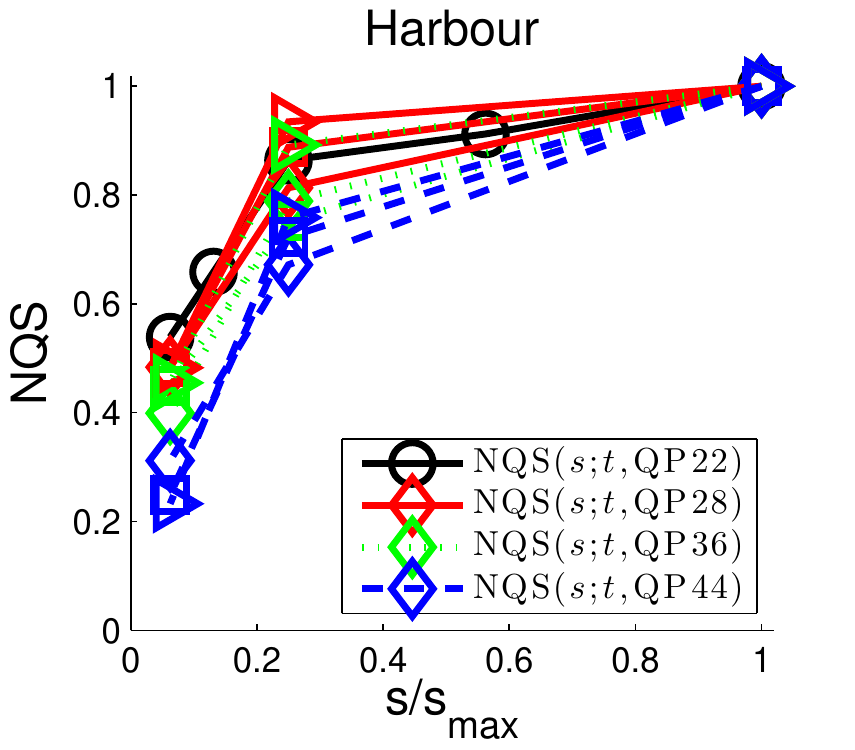}
\includegraphics[scale=0.5]{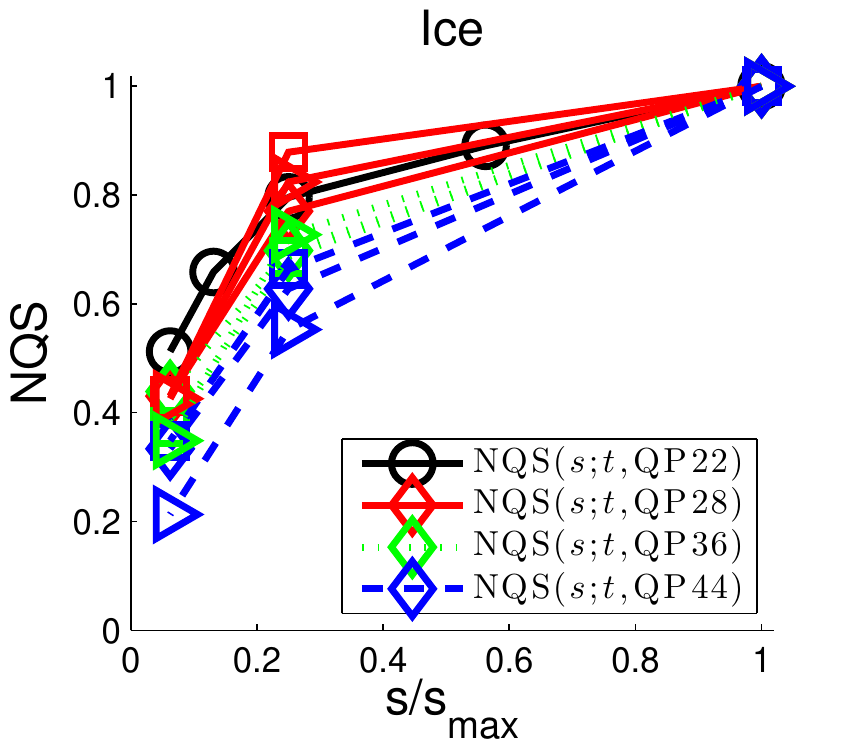}
\includegraphics[scale=0.5]{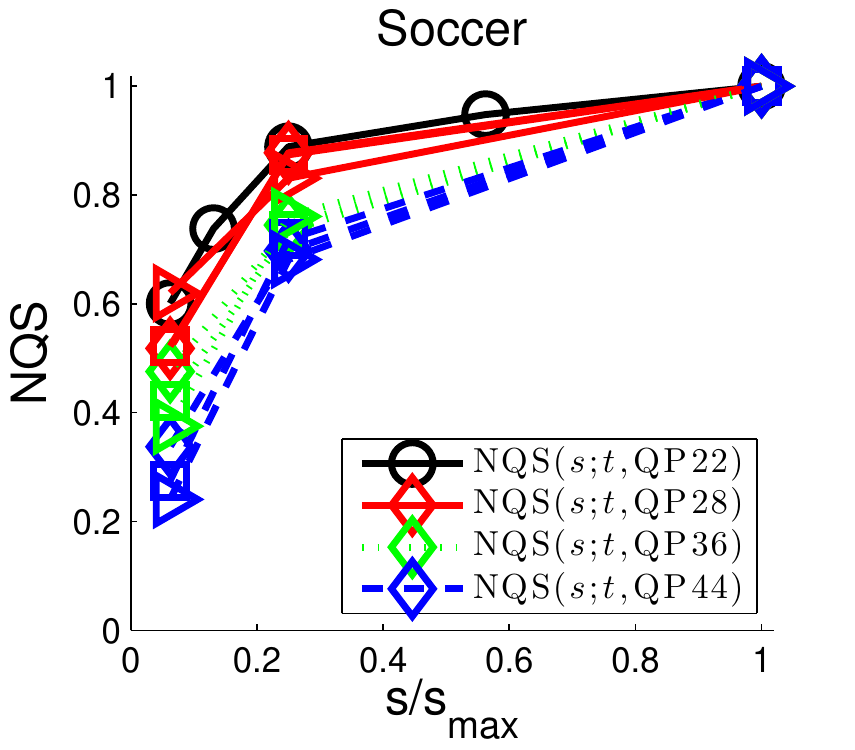}
\includegraphics[scale=0.5]{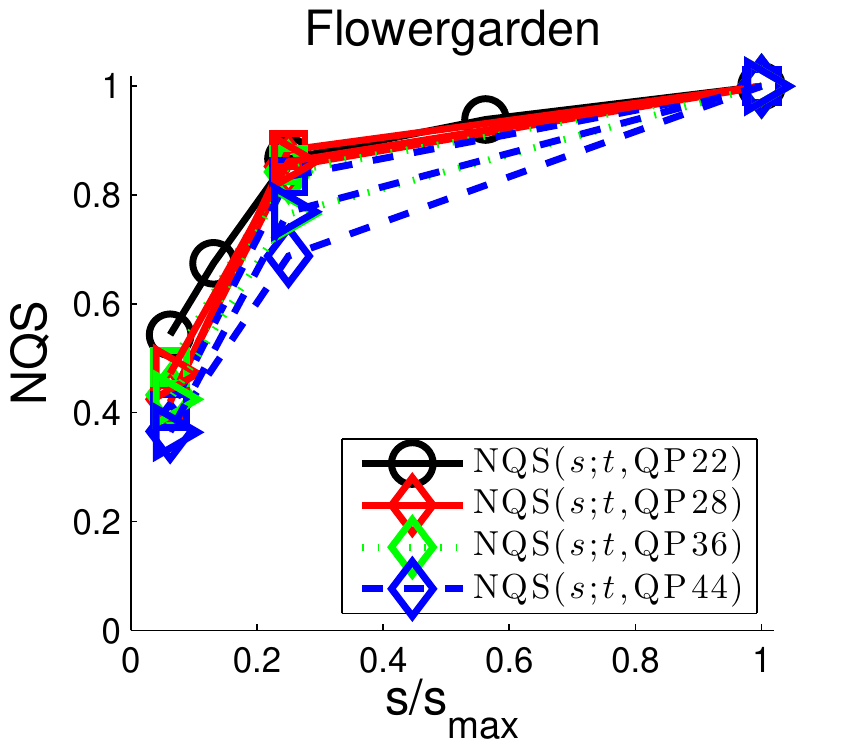}
\includegraphics[scale=0.5]{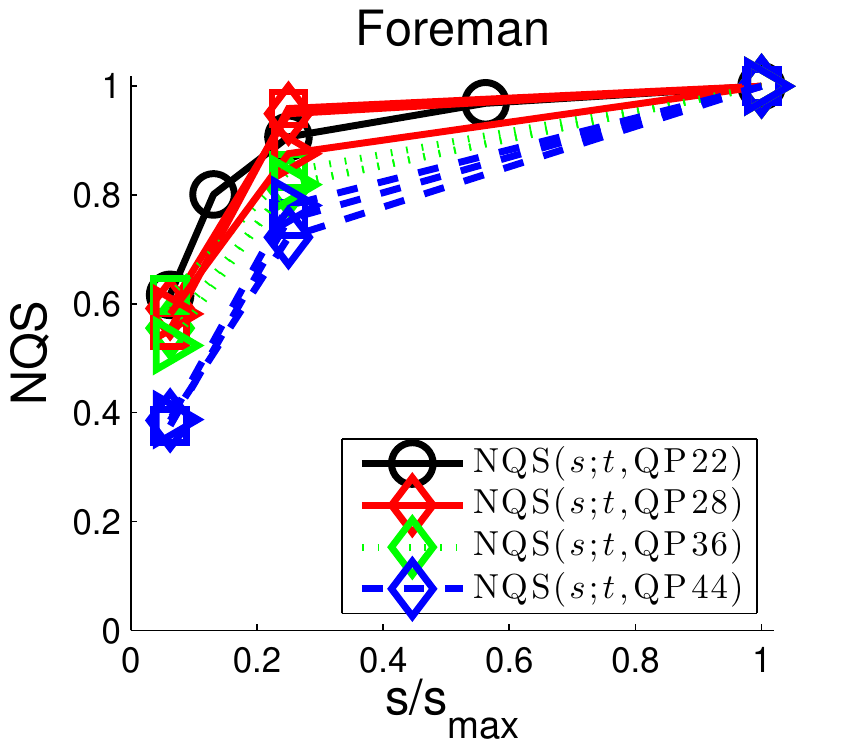}
\includegraphics[scale=0.5]{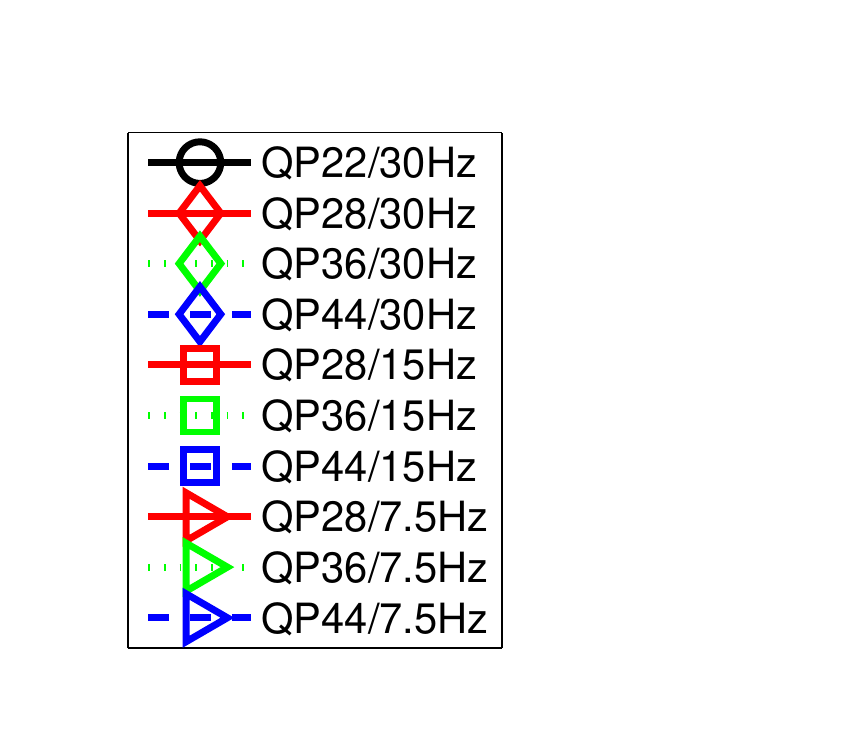}
\caption{Measured NQS under different QS's and TR's. Note that lines with the same color correspond to NQS data at different TR's but the same QP.}
\label{fig:NQS_data}
\end{figure*}
\begin{figure*}[!ht]
\vspace{-.in}
\centering
\includegraphics[scale=0.5]{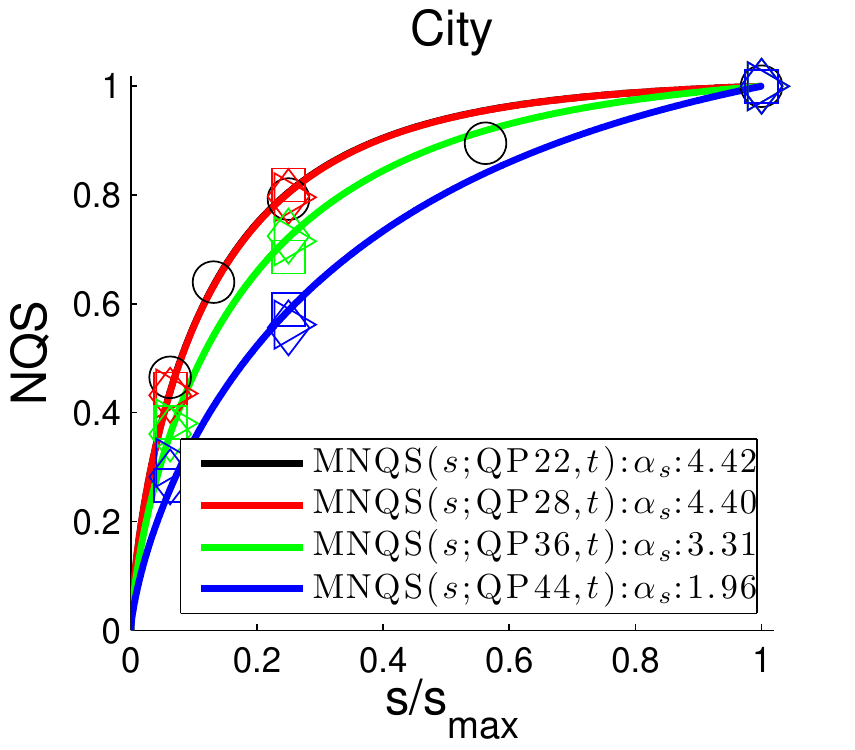}
\includegraphics[scale=0.5]{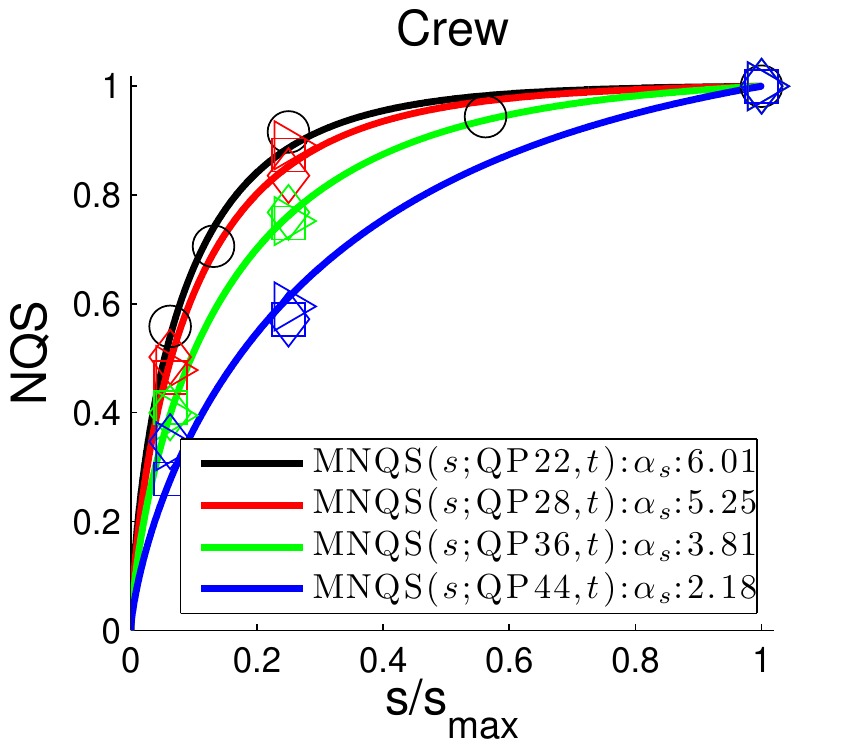}
\includegraphics[scale=0.5]{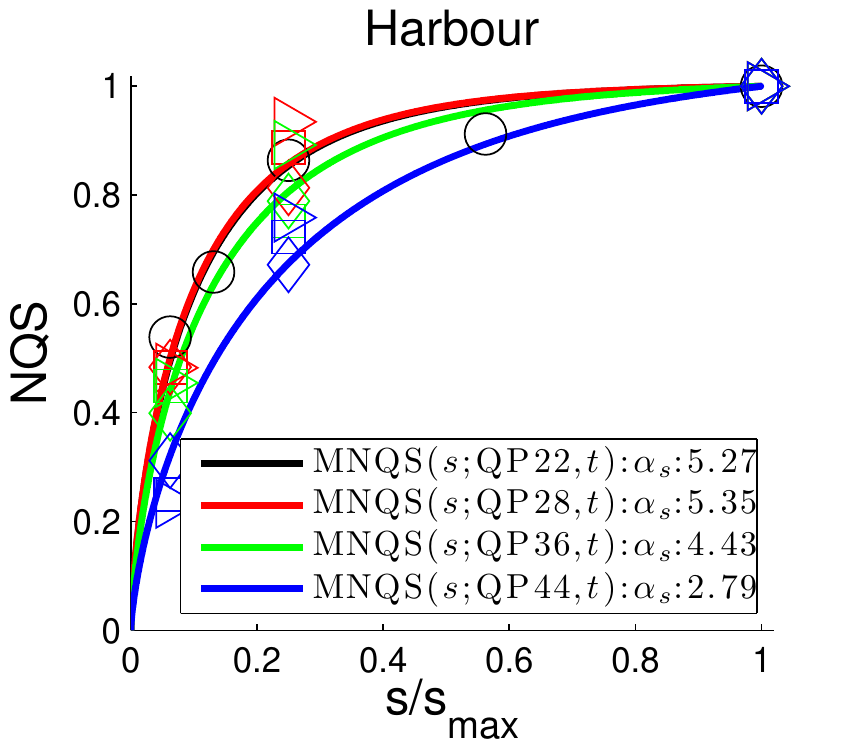}
\includegraphics[scale=0.5]{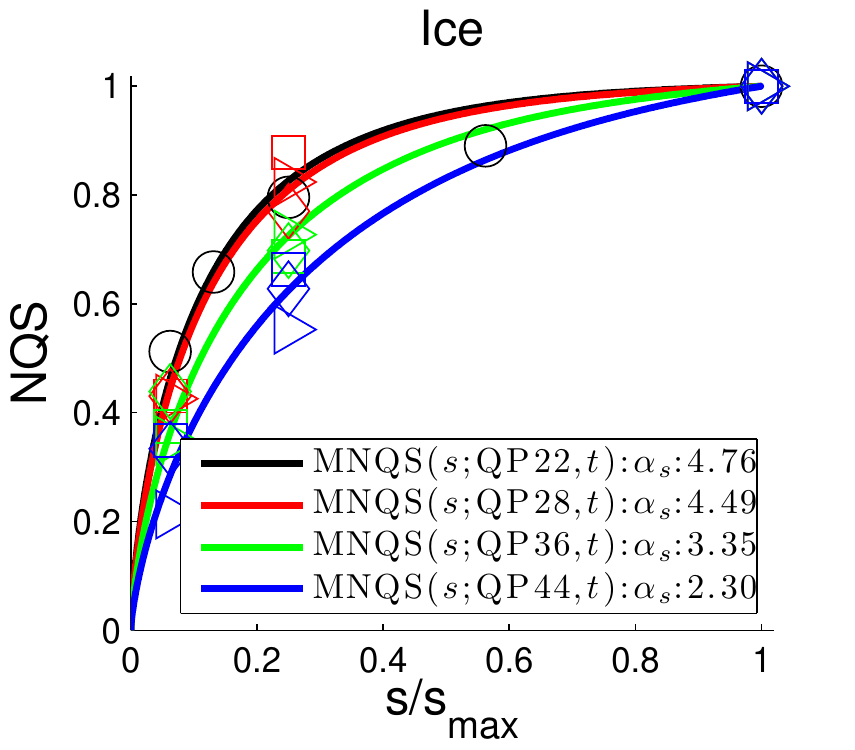}
\includegraphics[scale=0.5]{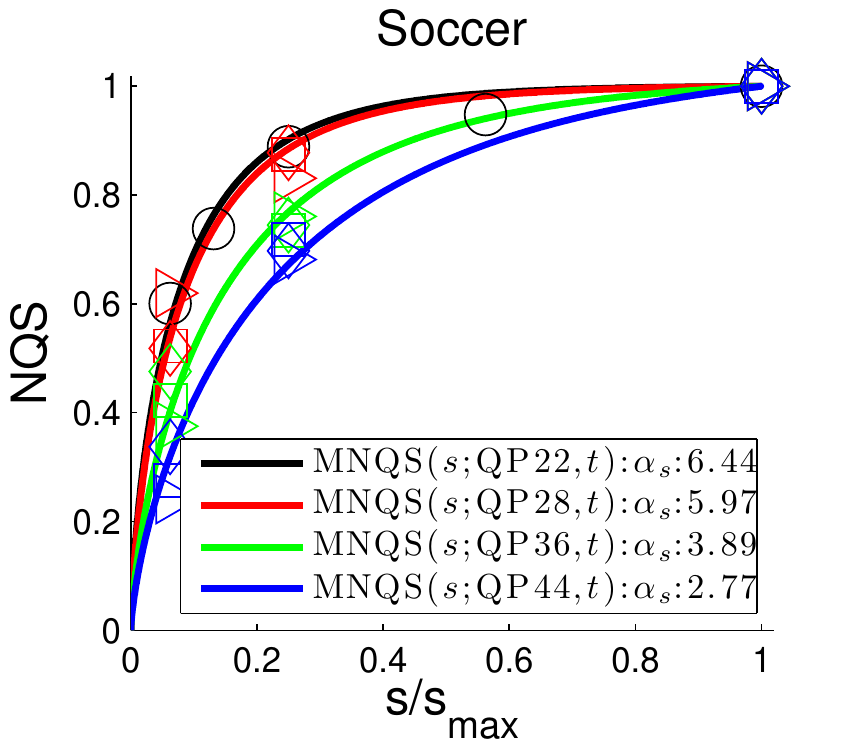}
\includegraphics[scale=0.5]{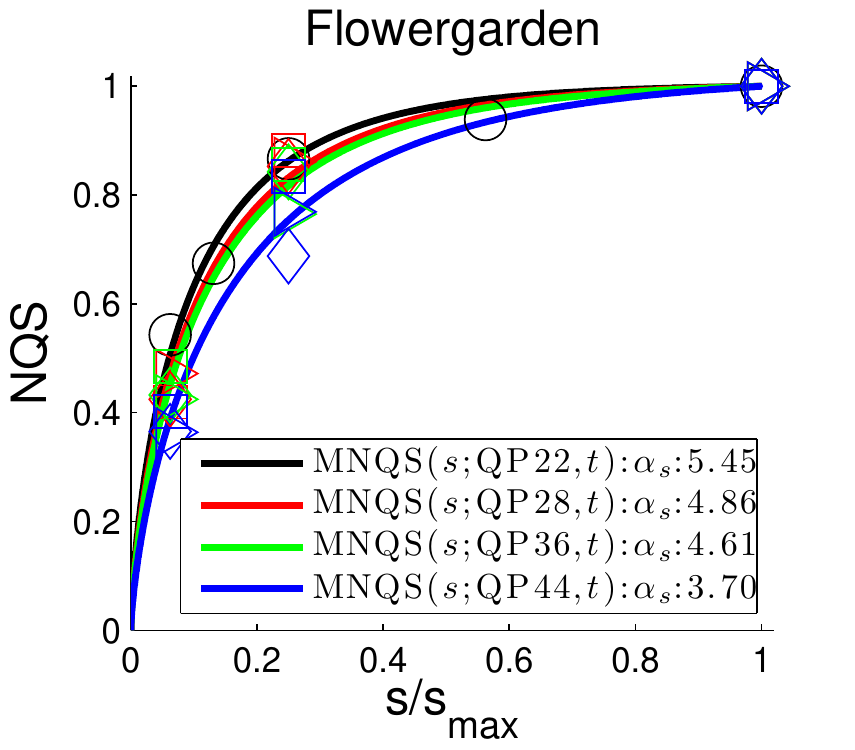}
\includegraphics[scale=0.5]{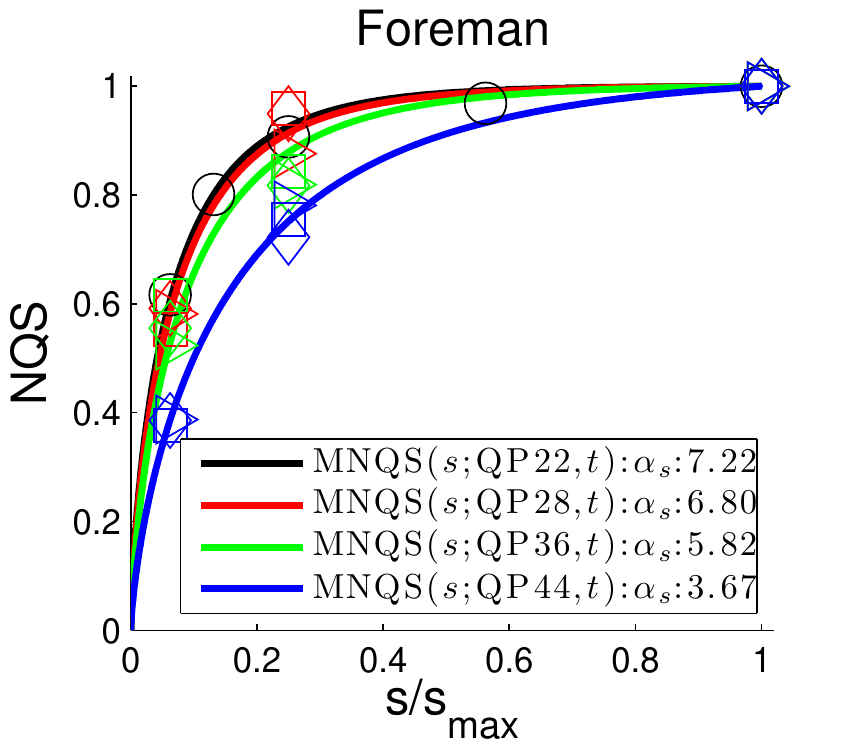}
\includegraphics[scale=0.5]{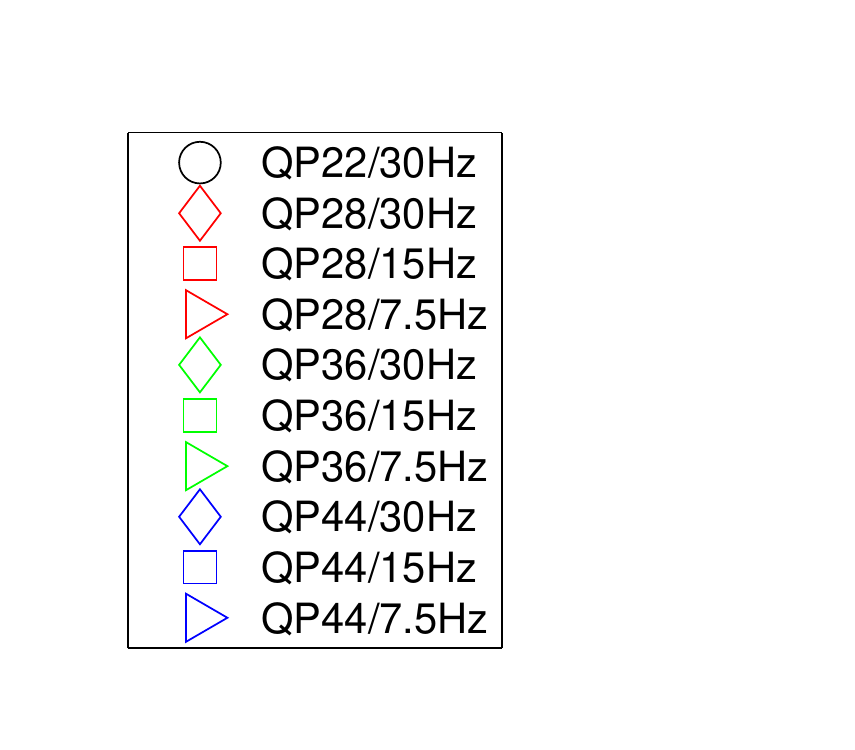}
\caption{Normalized quality v.s. normalized SR. Points are measured data under different QS's and TR's. Curves are derived from the model given in (\ref{eq:NQS_model}) with $\beta_s$=0.74. The parameter $\alpha_s$ for each sequence and QS is determined by least square fitting of data points at all TR's. PCC=$0.992$, RMSE=$0.03$.}
\label{fig:MNQS}
\end{figure*}

To map Test 1 data to Test 2, we use a single linear mapping function for all test sequences, because we only have one common PVS for each source sequence. To map Test 3 data to Test 2, since we have many common PVS's for each source video, we form a different linear mapping function for each video.

After combining, we scale the mapped Z-scores back to [0\ 10] scale, using:

    \begin{align}\label{eMOS}
    \vspace{-.25in}
    X_{mij,{\mathrm{scl}}} &= (\mathrm{MEDIAN}(\mathbb{X}^I_{\max}) - \mathrm{MEDIAN}(\mathbb{X}^I_{\min}))\nonumber \\
    & \times \frac{Z_{mij}-Z_{i,\min}}{Z_{i,\max} -Z_{i,\min}} + \mathrm{MEDIAN}(\mathbb{X}^I_{\min})),
    \end{align}
where $\mathrm{MEDIAN}(\cdot)$ represents the median operator. $\mathbb{X}^I_{\max}$ and $\mathbb{X}^I_{\min}$ are the set of all viewers' maximum and minimum ratings, respectively. $Z_{i,\max}$ and $Z_{i,\min}$ denote the maximum and minimum Z-scores of viewer $i$. With this scaling, the ratings from all viewers have a common range of $\mathrm{MEDIAN}(\mathbb{X}^I_{\min})$ to $\mathrm{MEDIAN}(\mathbb{X}^I_{\max})$.  In our subjective test data, $\mathrm{MEDIAN}(\mathbb{X}^I_{\min})=1$, and $\mathrm{MEDIAN}(\mathbb{X}^I_{\max})=10$.

 Finally, we average the scaled Z-scores from all viewers for each PVS to obtain its mean opinion score (MOS). The MOS for a sequence with a particular STAR combination, denoted by $s,t,q$, is indicated by ${\rm MOS}(s,t,q)$.



\section{Subjective Test Results and Proposed Quality Model}\label{sec:QSTAR}
In order to analyze the test results and derive a quality model reflecting the quality impact of SR, TR, and QS, we first explore how SR, TR or QS individually affects the quality ratings. In each of the following three subsections, we show how MOS varies with one variable (e.g., SR), while holding the other two variables fixed (e.g. TR and QS). Based on the trend observed from the data, we propose a mathematical model that characterizes the degradation of the quality with this variable (e.g. SR). We further examine the interactions of different variables through the two-way Analysis of Variance (ANOVA)~\cite{Snedecor_Statistics}. Finally in the last subsection, we propose an overall quality model by taking the product of the three model functions of individual variables, and validate its accuracy.

\subsection{Modeling Normalized Quality v.s. Spatial Resolution}\label{ssec:MNQS}
In this subsection, we examine how SR affects the perceived quality, when TR and QS are fixed. Towards this goal, we plot the normalized quality v.s. normalized SR $s/s_{\max}$ (NQS) (here, $s_{\max}$ = 4CIF) at the same TR and QS in Fig.~\ref{fig:NQS_data} for each source sequence. The NQS function is defined as
\begin{align}\label{eq:NQS}
{\rm NQS}(s;t,q)=\frac{\textrm{MOS}(s,t,q)}{\textrm{MOS}(s_{\max},t,q)}.
\end{align}
From Fig.~\ref{fig:NQS_data}, we can observe that the dropping curves of different TR's but same QS tend to cluster together except for a few sequences (\emph{Flowergarden}, \emph{Harbour} and \emph{Ice}) at high QS (QP=44, or QS=102).  On the other hand, the dropping trend in all cases seem to depend on the QS, with a higher QS leading to a faster dropping, except for the two lowest QS's, which lead to similar dropping rates.
To examine the dependency of the NQS on TR and QS, respectively, we conduct a three-way ANOVA test for NQS data, which is described later in Sec.~\ref{sec:ANOVA_analysis}.

\begin{figure}[htp]
\vspace{-.in}
\centering
\includegraphics[scale=0.63]{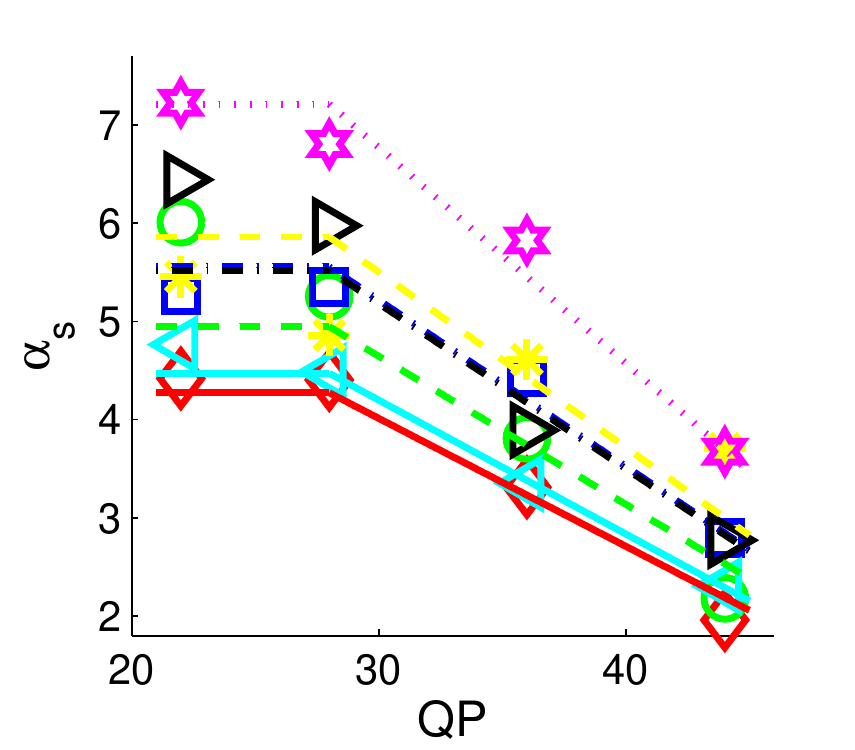}
\includegraphics[scale=0.3]{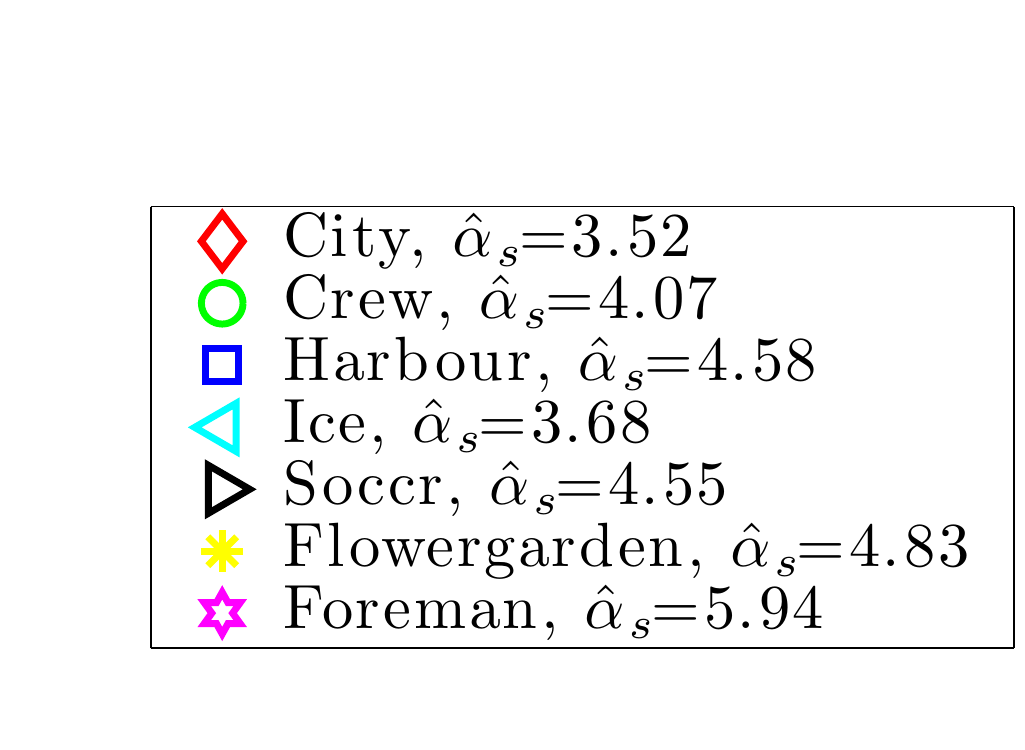}
\caption{The Predicted (in curves) and original (in points) $\alpha_s$ v.s. QP for each sequence.}
\label{fig:alpha_s_QP}
\end{figure}

Based both on our observation of the dependency of NQS dropping rate on QS and TR, and on the results of the ANOVA test on NQS in Tab.~\ref{tab:ANOVA_dependency_test}, we choose to approximate the NQS data at the same QS but different TR's with the same model function.  By examining the general trend of how NQS changes with normalized SR, we propose the following model, called MNQS, i.e.,
\begin{equation}\label{eq:NQS_model}
{\rm MNQS}(s;q)=\frac{1- e^{-\alpha_s(q)(\frac{s}{s_{\max}})^{\beta_s}}}{1-e^{-\alpha_s(q)}},
\end{equation}
where $\alpha_s(q)$ characterizes the quality decay rate as $s$ decreases, with a smaller value corresponding to a faster dropping rate.  The parameter $\beta_s$ controls the general shape of the inverse exponential function given in (\ref{eq:NQS_model}), and is called the shaping parameter, so that only a single parameter $\alpha_s$ is content- and QS-dependent. Figure~\ref{fig:MNQS} shows that this model fits the NQS data at different QP's very well. To quantify the accuracy of the fitting, we measure the Pearson Correlation Coefficient (PCC) and root mean square error (RMSE) between the measured and predicted data, which are given in Fig.~\ref{fig:MNQS}. The parameter $\alpha_s$ for each QP is obtained by least squares fitting to NQS data at this QP but all different TR's.
We further examine the model performance when we allow $\alpha_s$ to vary with TR. Table~\ref{tab:bt_bs_dependency} shows that this does not lead to significant improvement in PCC and RMSE. This further confirms that by assuming $\alpha_s$ to be independent of TR, we can reduce the model complexity without sacrificing the model accuracy.

\begin{figure*}[htp]
\vspace{-.in}
\centering
\includegraphics[scale=0.5]{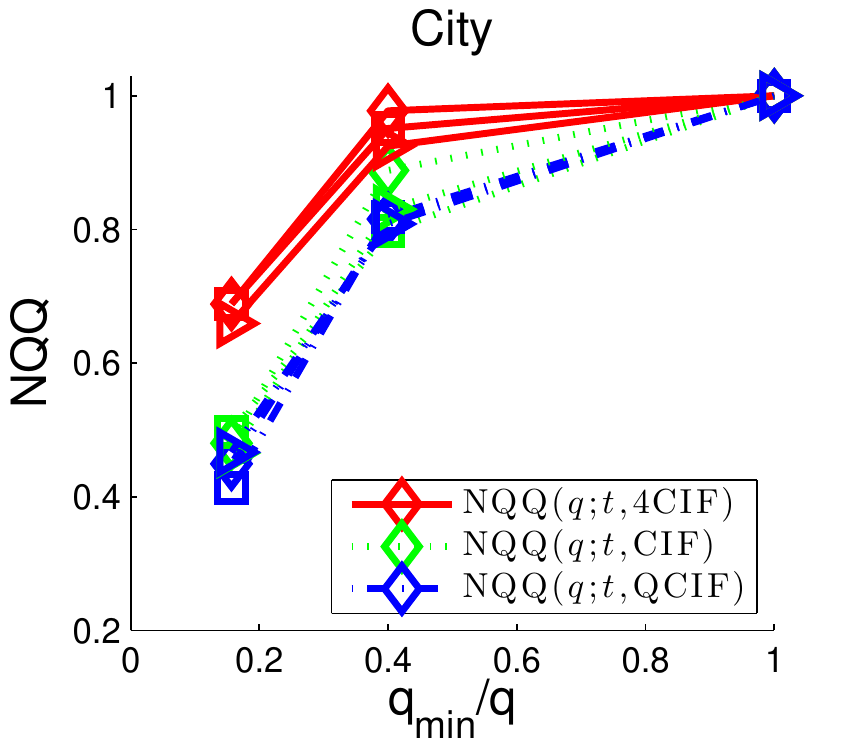}
\includegraphics[scale=0.5]{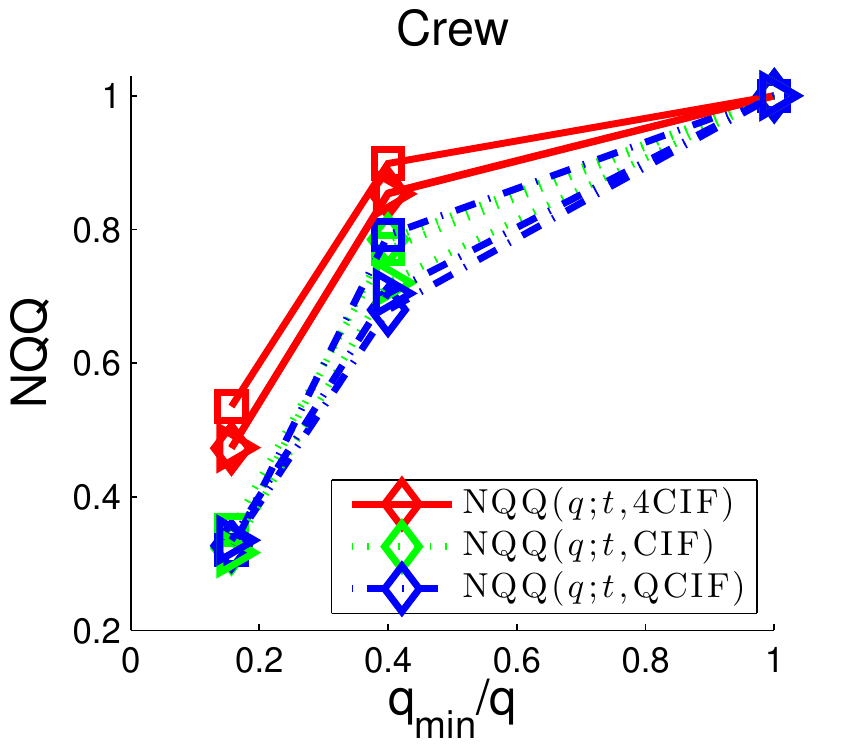}
\includegraphics[scale=0.5]{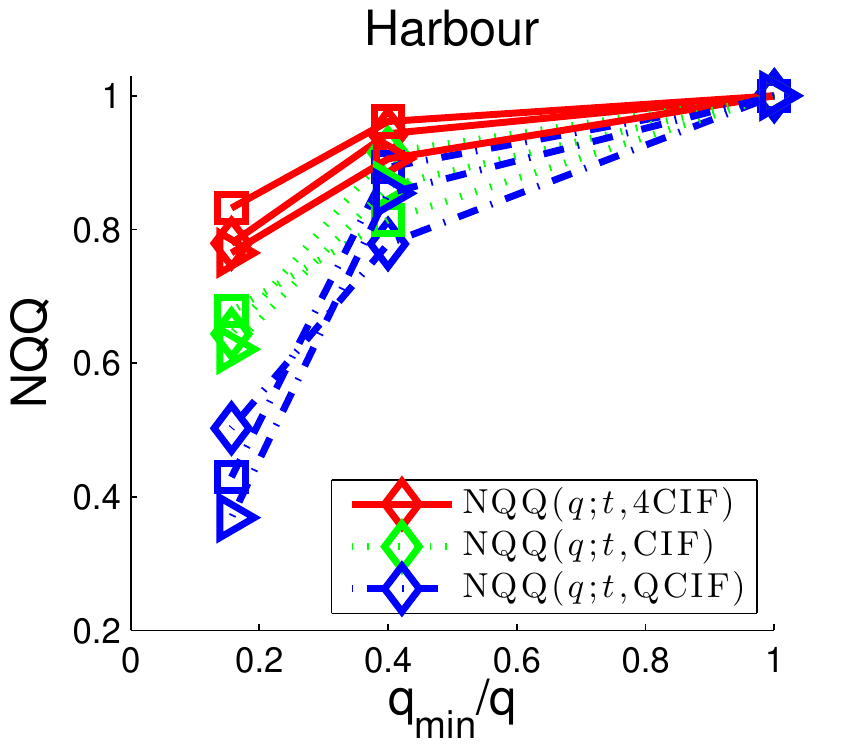}
\includegraphics[scale=0.5]{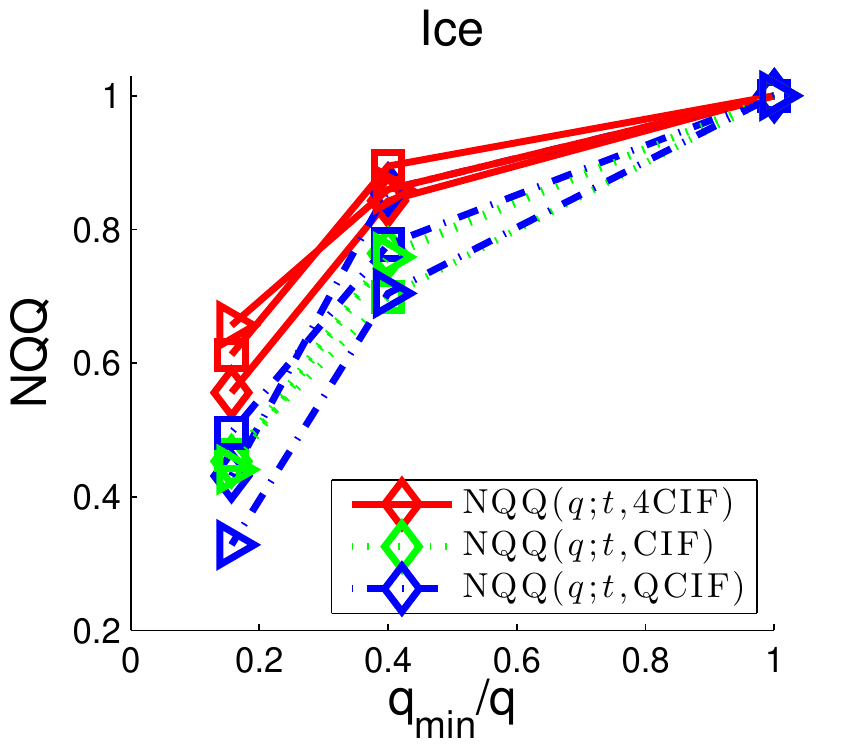}
\includegraphics[scale=0.5]{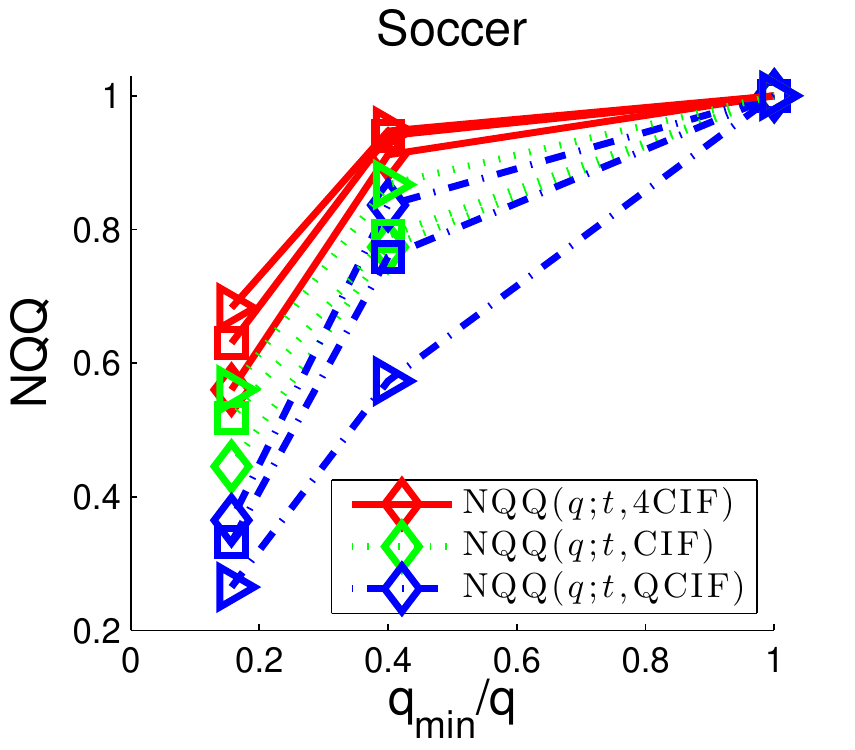}
\includegraphics[scale=0.5]{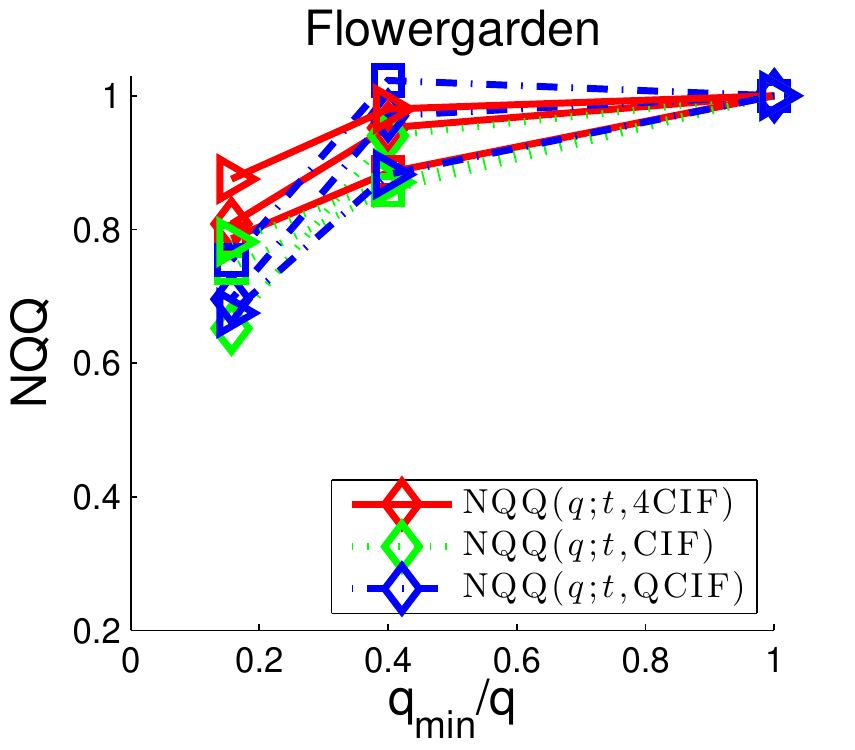}
\includegraphics[scale=0.5]{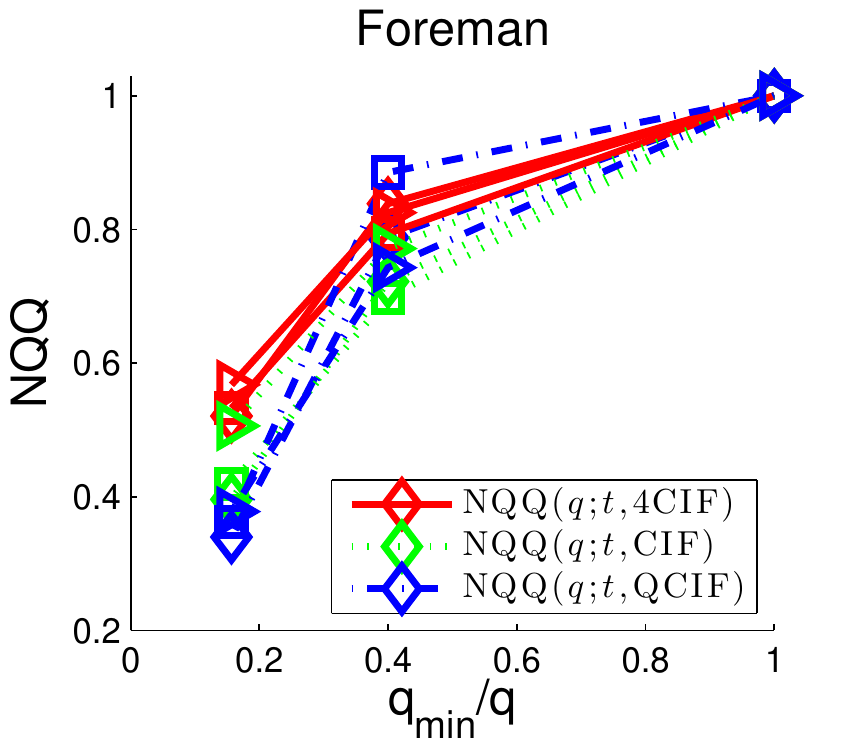}
\includegraphics[scale=0.5]{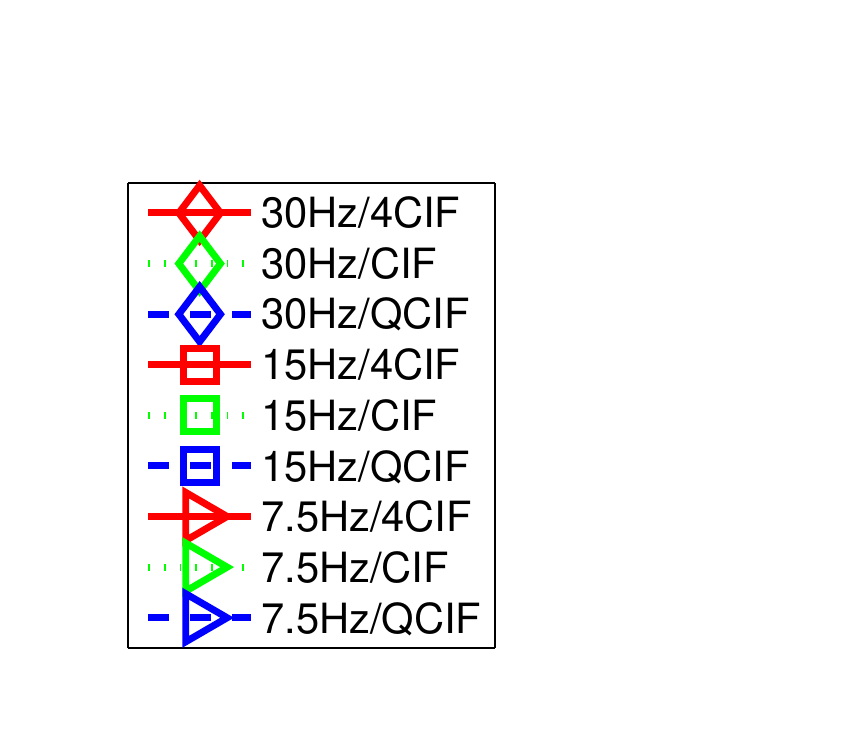}
\caption{Measured NQQ under different SR's and TR's. Note that lines with the same color correspond to NQQ data at different TR's but the same SR.}
\label{fig:NQQ_data}
\end{figure*}

%
\begin{figure*}[!htp]
\centering
\includegraphics[scale=0.5]{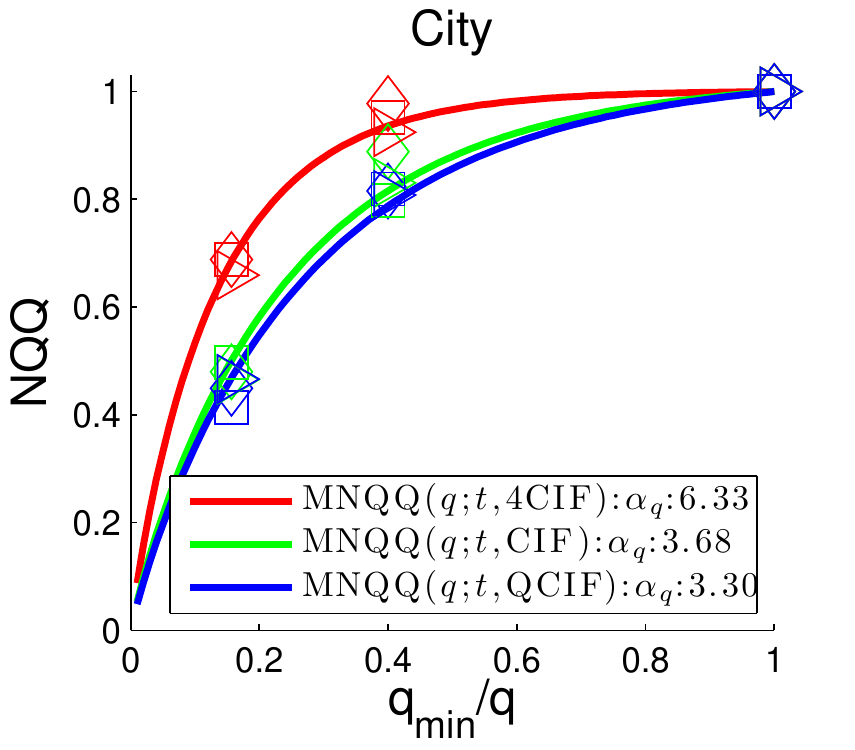}
\includegraphics[scale=0.5]{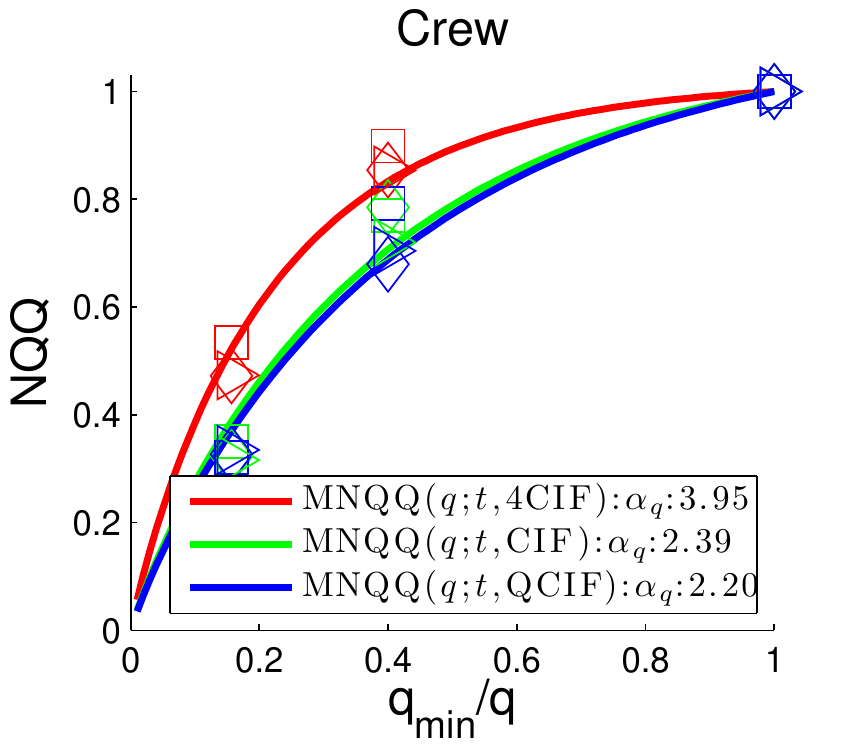}
\includegraphics[scale=0.5]{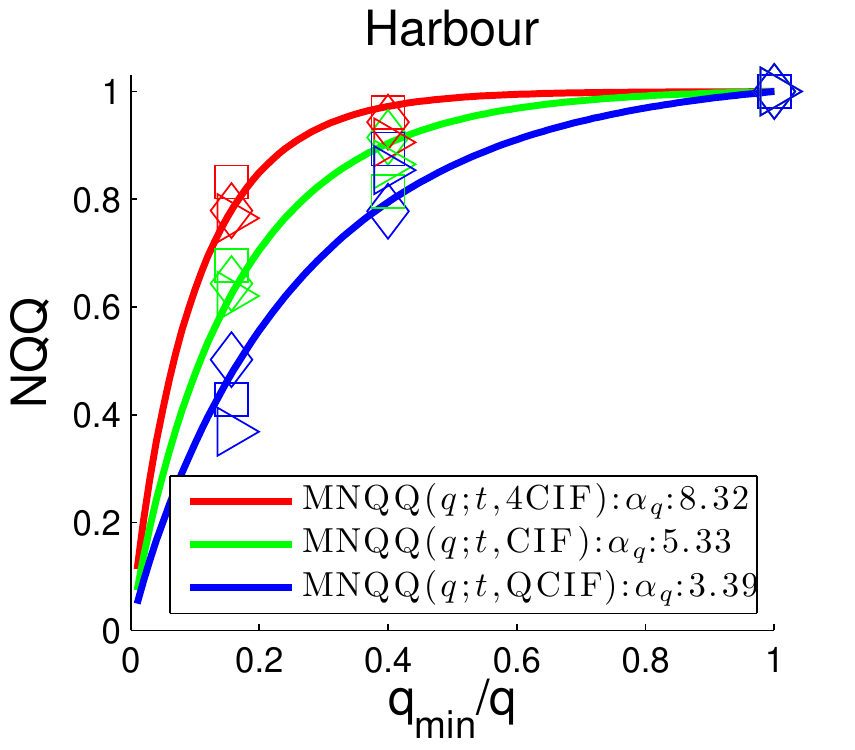}
\includegraphics[scale=0.5]{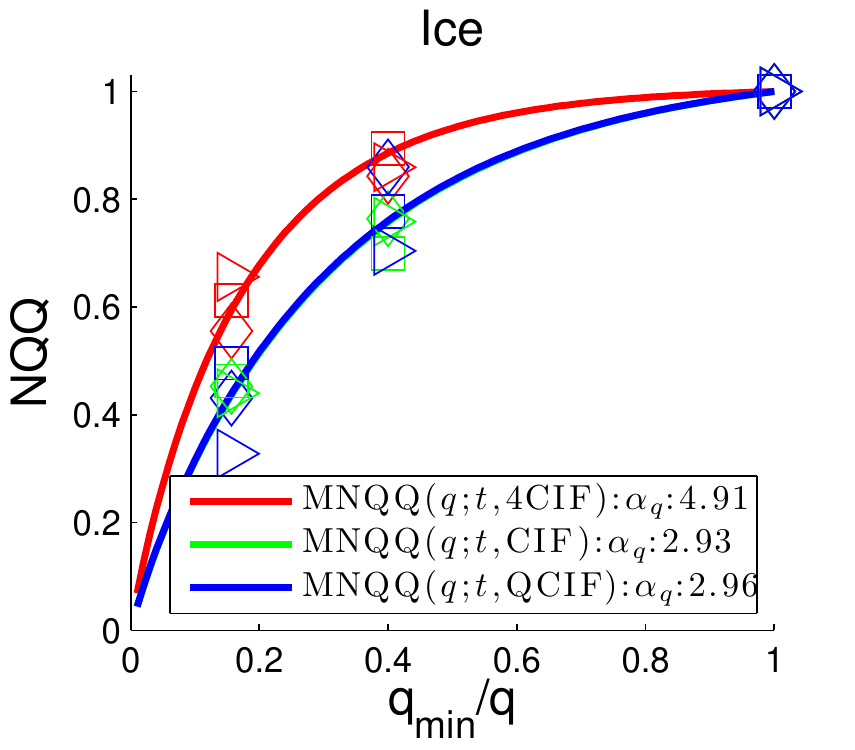}
\includegraphics[scale=0.5]{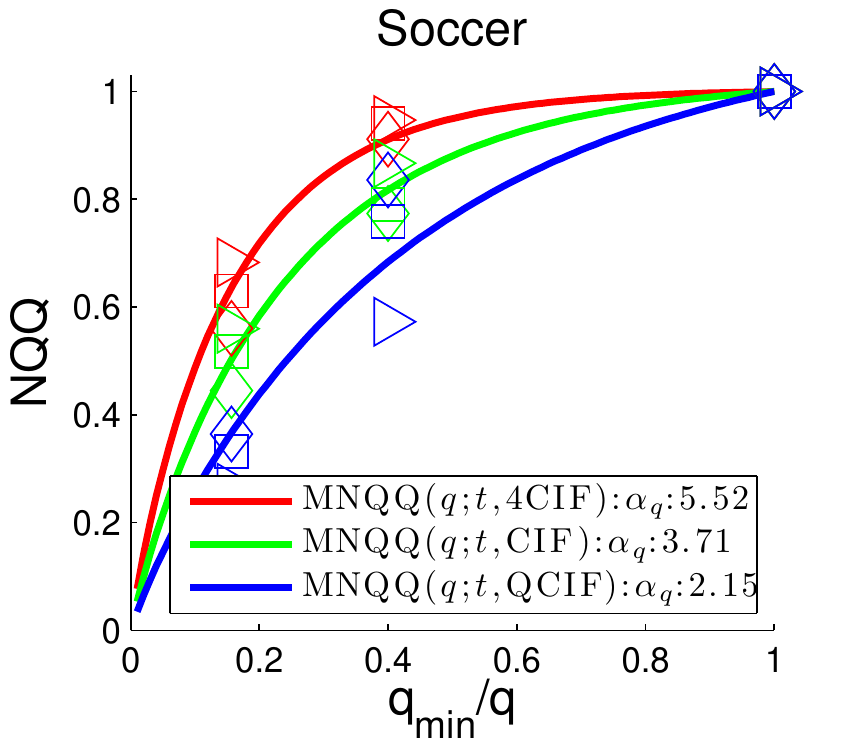}
\includegraphics[scale=0.5]{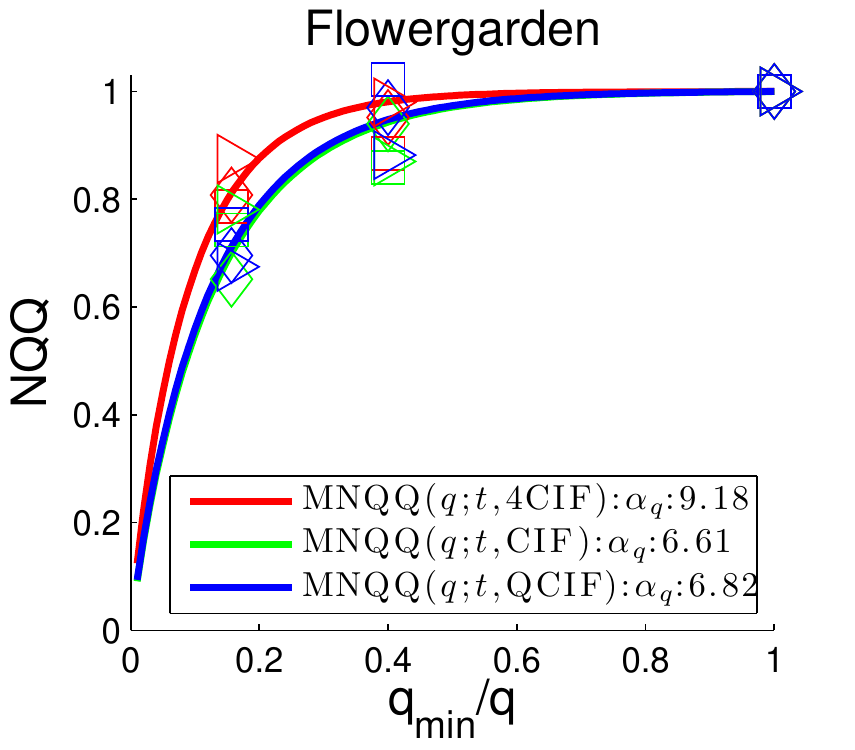}
\includegraphics[scale=0.5]{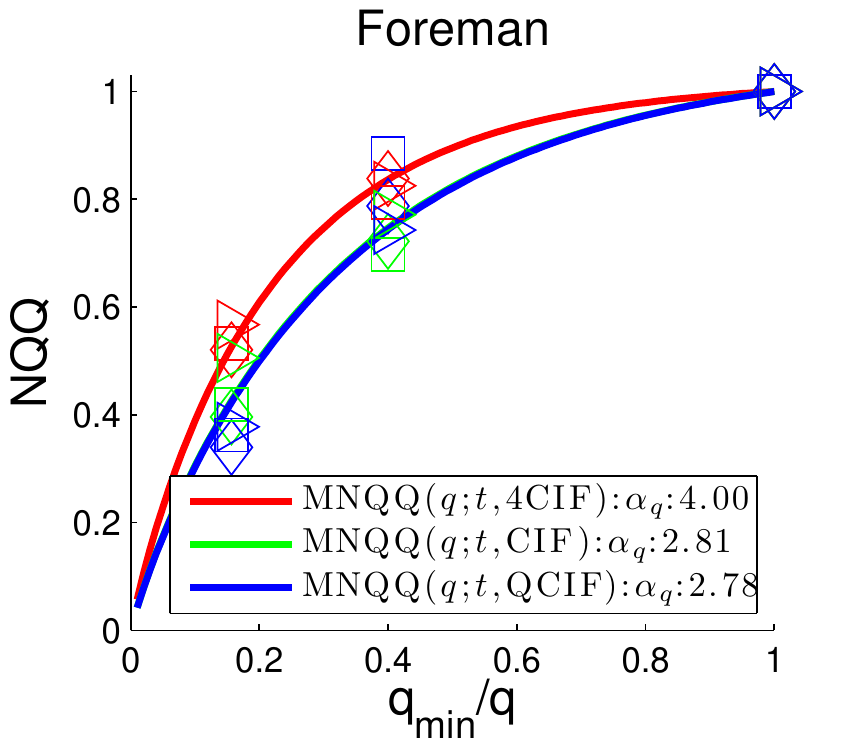}
\includegraphics[scale=0.5]{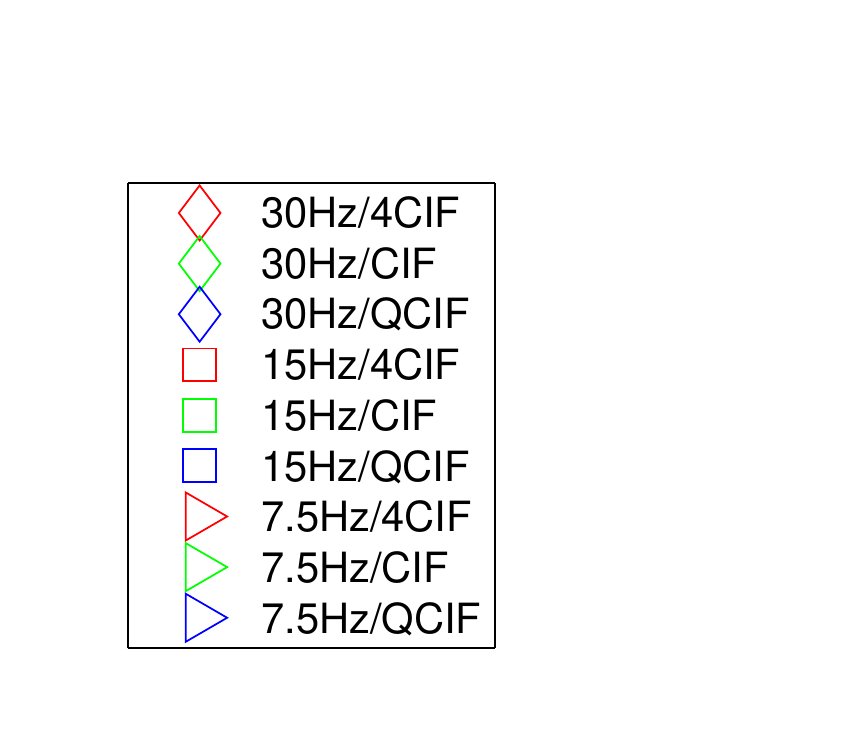}
 \caption{Normalized  quality v.s. inversed normalized QS. Points are measured data under different SR's and TR's. Curves are derived from the model in  (\ref{eq:NQQ_model}) for all $t$ at a given $s$. The parameter $\alpha_q$ for each sequences and SR is determined by least squares fitting of data at all TR's. PCC=$0.982$, RMSE=$0.041$.}\label{fig:MNQQ_model_invexp}
\end{figure*}



In Fig.~\ref{fig:MNQS} each subplot contains the MNQS curves corresponding to different QS's, for the same video content. We can see that the quality drops faster at larger QS. This is because larger QS introduces more blurring artifacts\footnote{Note that with the use of deblocking filter in H.264 encoder, quantization artifacts at high QP manifest as blurring.} compared with smaller QS given the same SR.

To further simplify the model, we investigate the relationship between $\alpha_s$ and QS. Figure~\ref{fig:alpha_s_QP} shows that $\alpha_s$ has an approximately linear relationship with QP, for QP $>=$28, and the $\alpha_s$ for QP=22 is very close to that for QP=28. Therefore, we propose to model the dependency of $\alpha_s$ on $q$ (and equivalently on QP) by
\begin{align}\label{eq:alpha_s_QP}
\alpha_s(q) &= \hat{\alpha}_s L({\rm QP}(q)),\nonumber\\
\mbox{~with~}L({\rm QP}) &= \left\{ \begin{array}{ll}
\centering
\upsilon_1{\rm QP}+\upsilon_2, & \textrm{if QP} >= 28\\
28 \upsilon_1+\upsilon_2,  & \textrm{if QP} < 28,
\end{array} \right.
\end{align}
where QP is related to $q$ with ${\rm QP}(q) = 4+6\log_2q$, as defined by the H.264/SVC codec~\cite{H264SVC}. We derive the constants $\upsilon_1$, $\upsilon_2$, and $\beta_s$ (which are sequence independent) together with the model parameter $\hat{\alpha}_s$ (sequence dependent) by minimizing the mean squares error between the measured NQS data at all STAR combinations and the predicted NQS using (\ref{eq:NQS_model}) and (\ref{eq:alpha_s_QP}). The best fitting constants are $\upsilon_1=-0.037$, $\upsilon_2=2.25$, and $\beta_s=0.74$.
%
Figure~\ref{fig:alpha_s_QP} shows that the $\alpha_s$ determined using (\ref{eq:alpha_s_QP}) are quite close to the original $\alpha_s$, except for a few cases (e.g. \emph{Flowergarden} and \emph{Soccer}). Even in those cases, the differences in $\alpha_s$ values do not have a significant impact on the resulting MNQS curves. The MNQS curves obtained using (\ref{eq:NQS_model}) and (\ref{eq:alpha_s_QP}) with only a single content-dependent parameter $\hat{\alpha}_s$ are very similar to those shown previously in Fig.~\ref{fig:MNQS}, and hence are not included to save the space.  Table~\ref{tab:bt_bs_dependency} shows that using a single parameter $\hat{\alpha}_s$ is only slightly worse than using independently determined $\alpha_s$ for each QP.
Therefore, we propose to use~(\ref{eq:alpha_s_QP}) together with~(\ref{eq:NQS_model}) to model NQS, which needs only one content-dependent parameter $\hat{\alpha}_s$ across different QP levels.
\begin{figure*}[htp]
\vspace{-.in}
\centering
\includegraphics[scale=0.5]{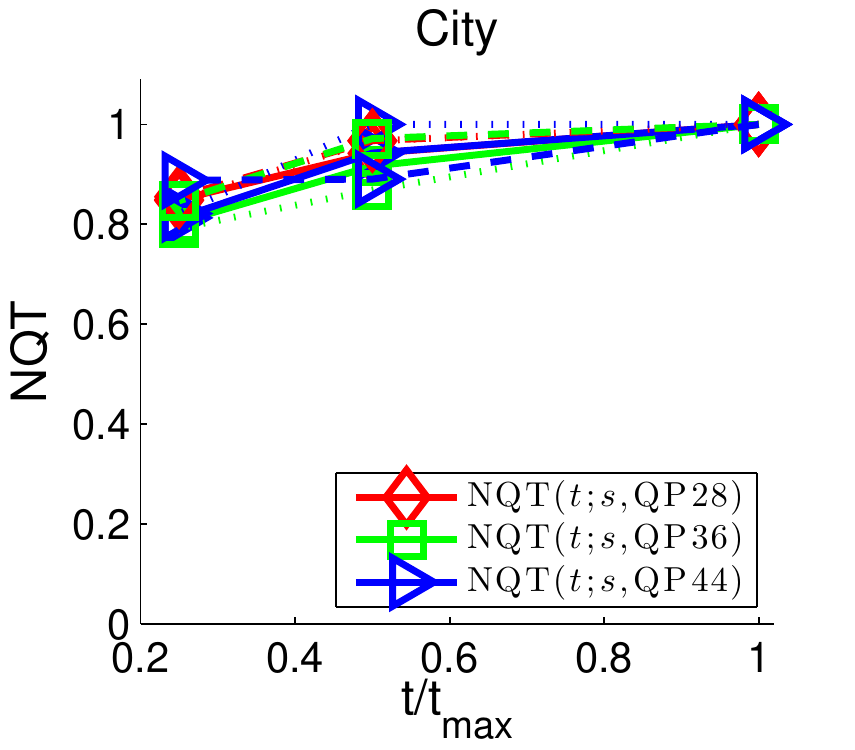}
\includegraphics[scale=0.5]{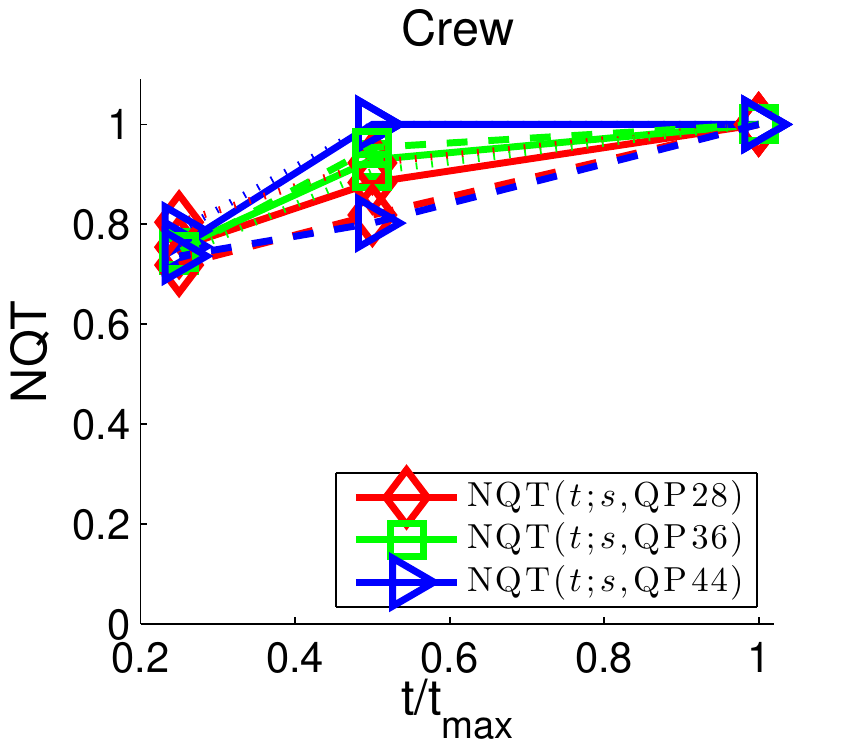}
\includegraphics[scale=0.5]{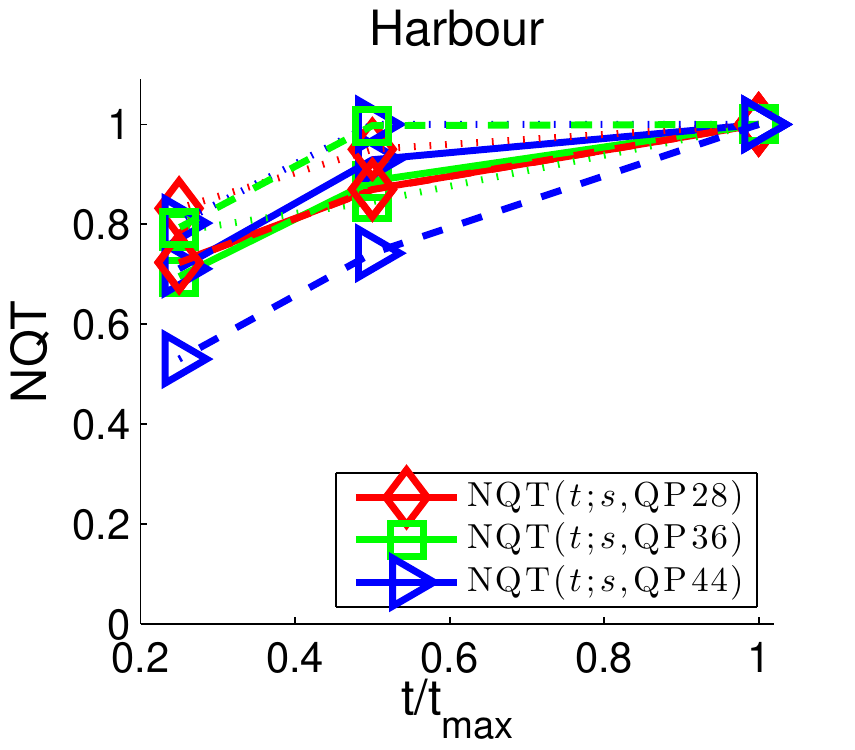}
\includegraphics[scale=0.5]{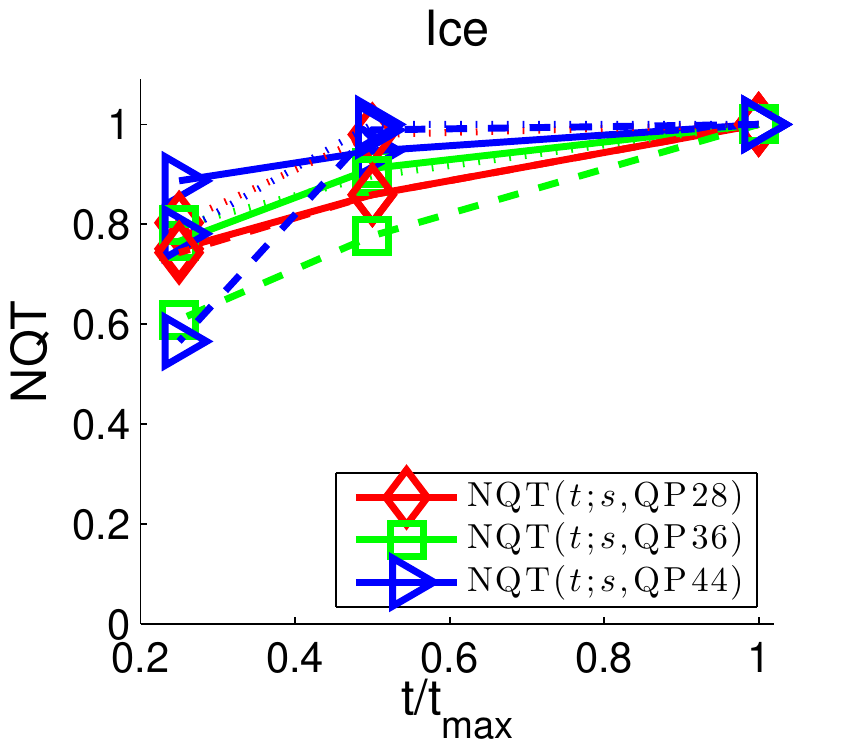}
\includegraphics[scale=0.5]{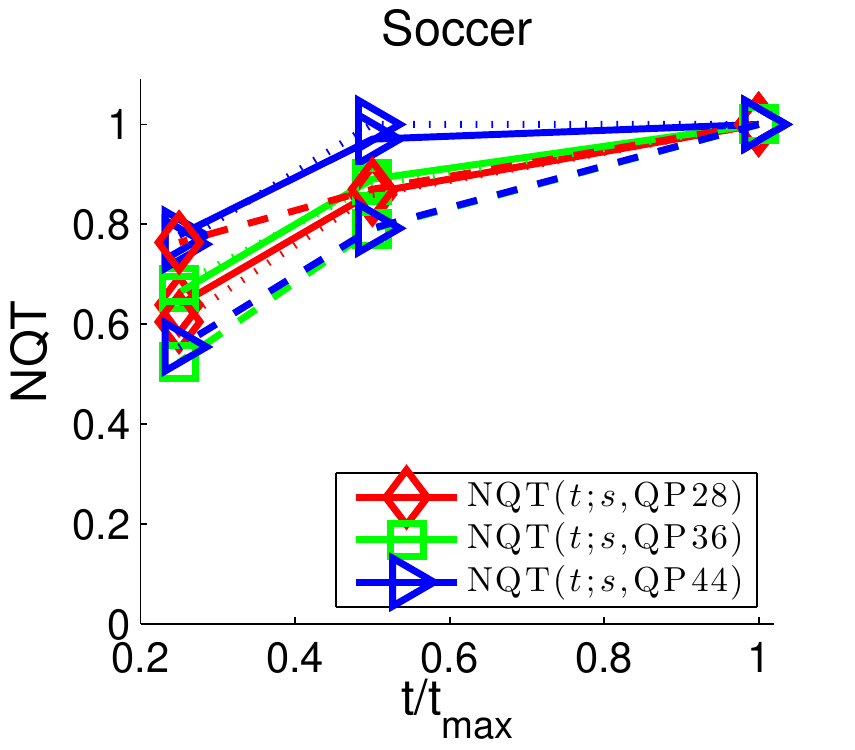}
\includegraphics[scale=0.5]{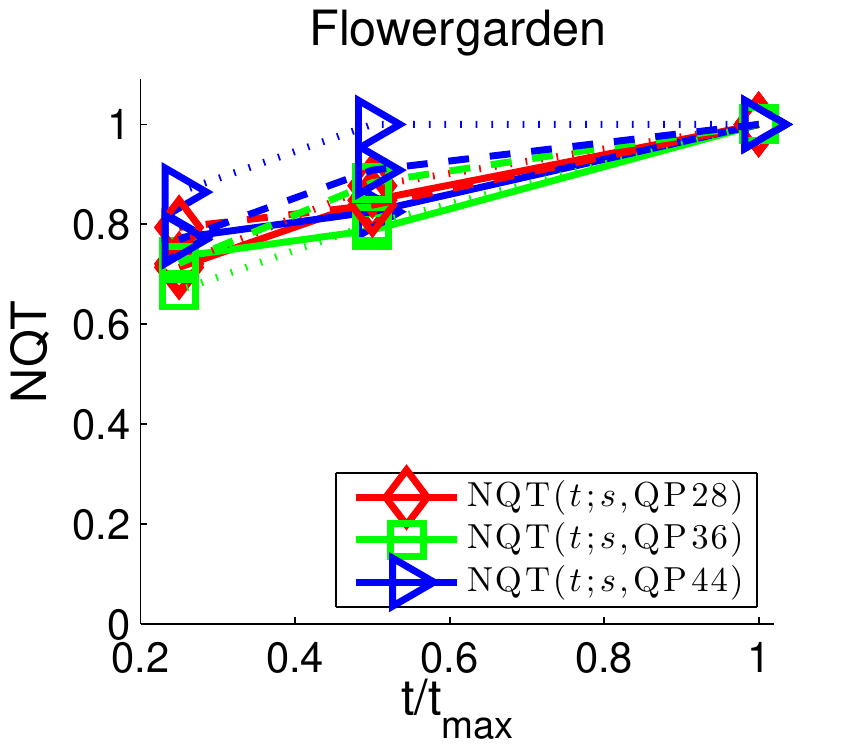}
\includegraphics[scale=0.5]{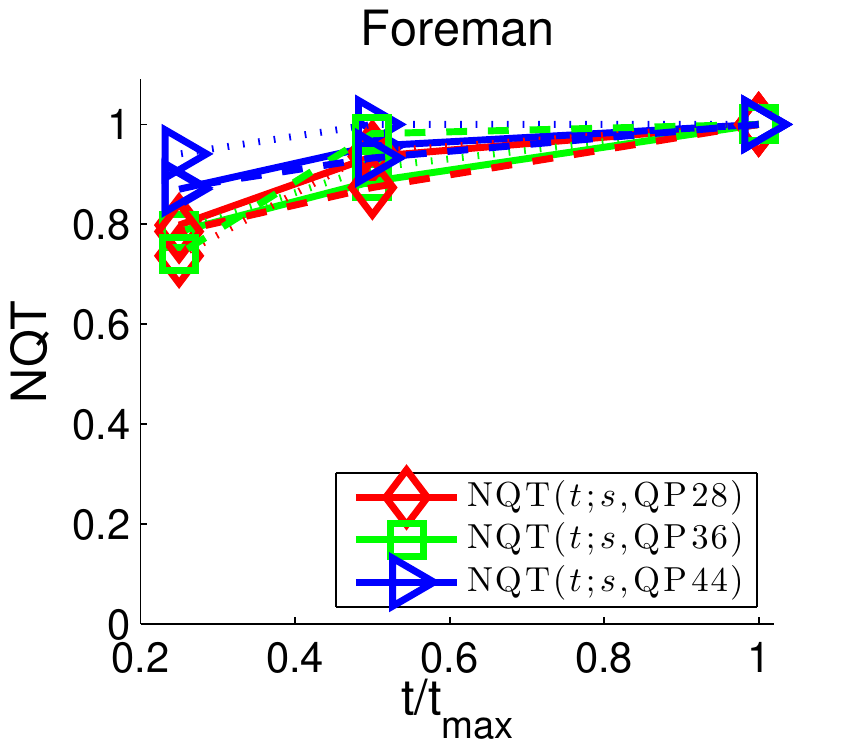}
\includegraphics[scale=0.5]{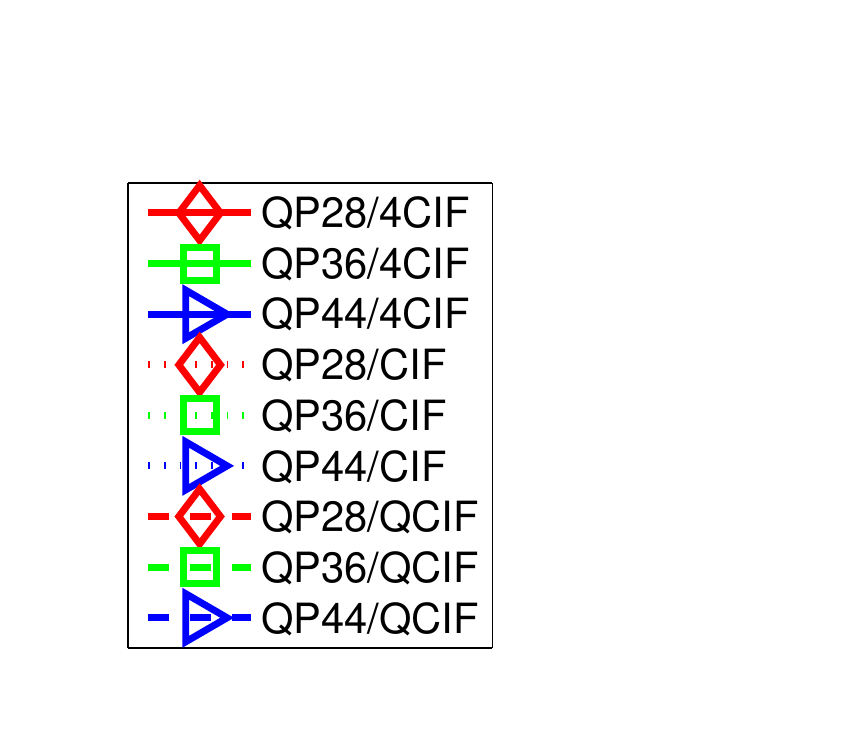}
\caption{Measured NQT under different QS's and SR's. Note that lines with the same color correspond to NQT data for test sequences at different SR's (4CIF, CIF, QCIF),  but the same QP.}
\label{fig:NQT_data}
\end{figure*}
\begin{figure*}[htp]
\vspace{-. in}
\centering
\includegraphics[scale=0.5]{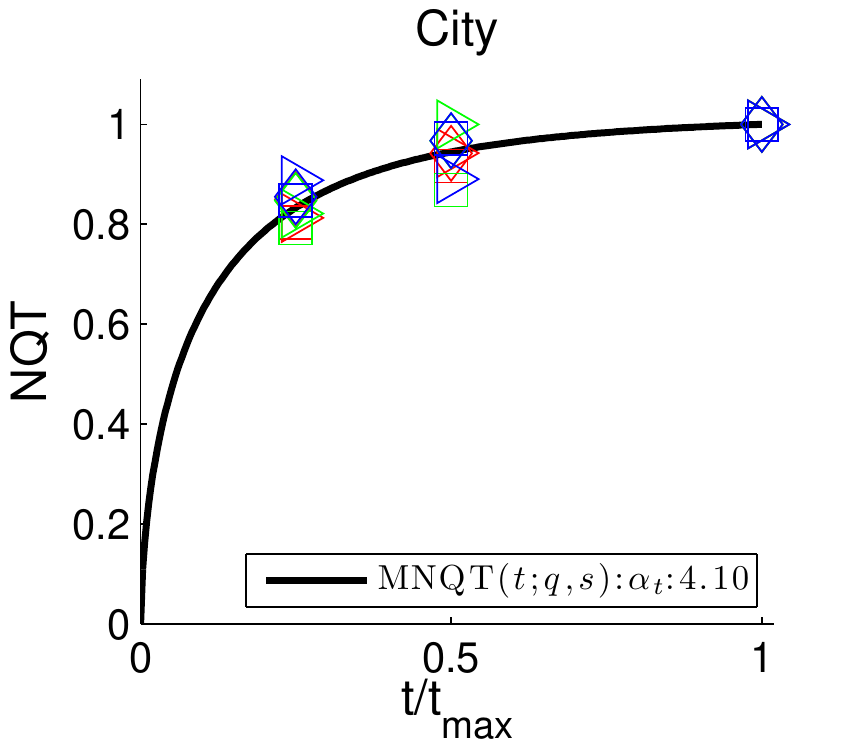}
\includegraphics[scale=0.5]{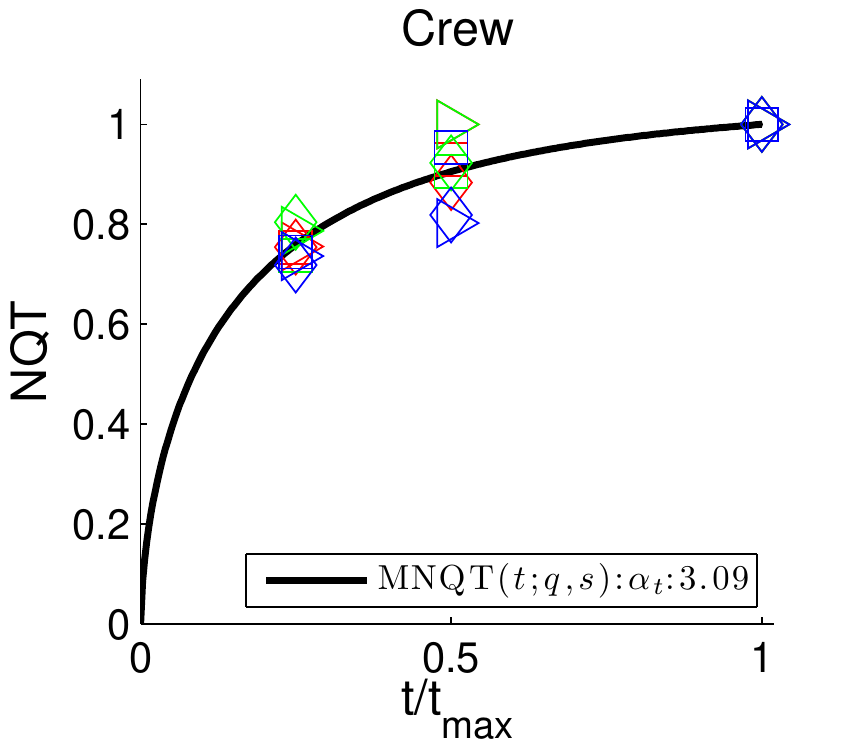}
\includegraphics[scale=0.5]{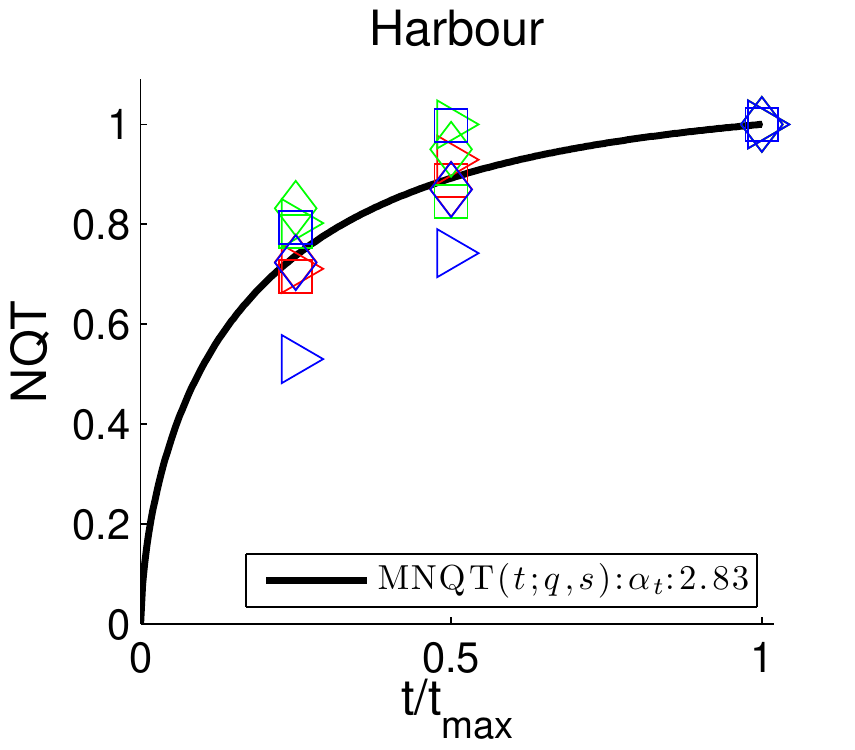}
\includegraphics[scale=0.5]{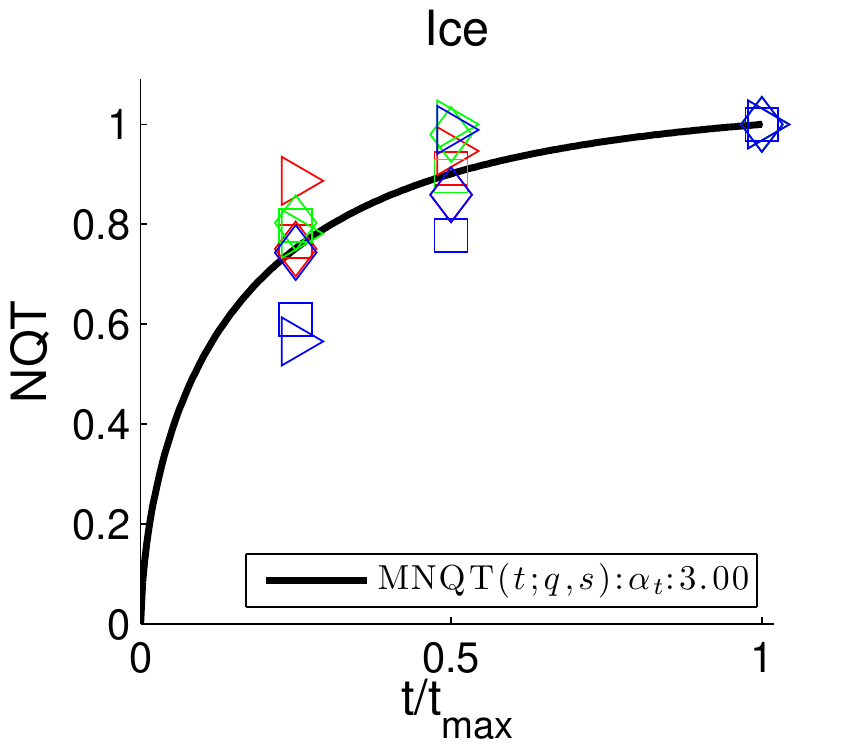}
\includegraphics[scale=0.5]{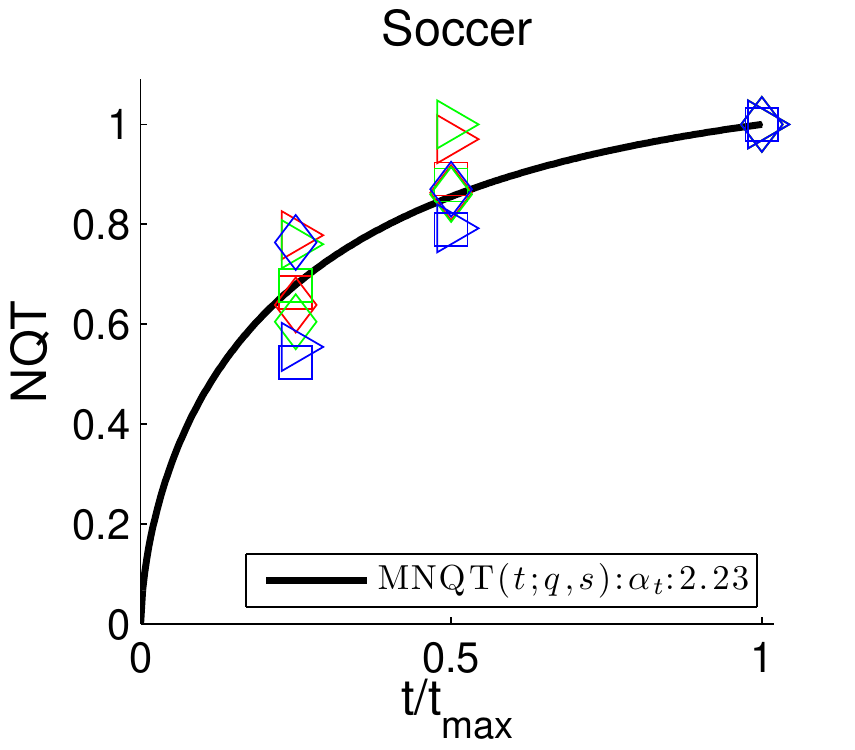}
\includegraphics[scale=0.5]{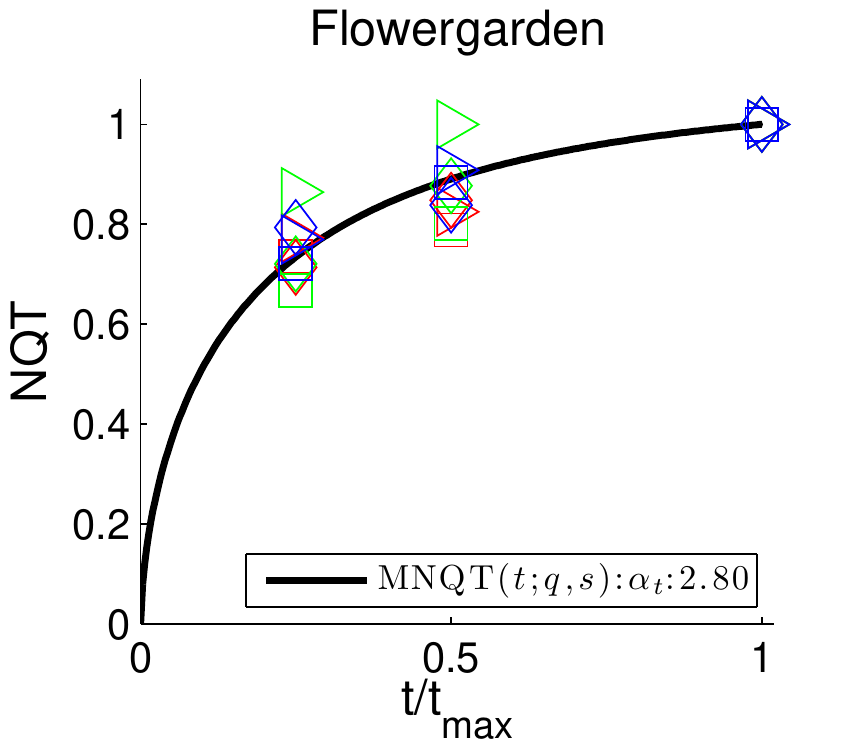}
\includegraphics[scale=0.5]{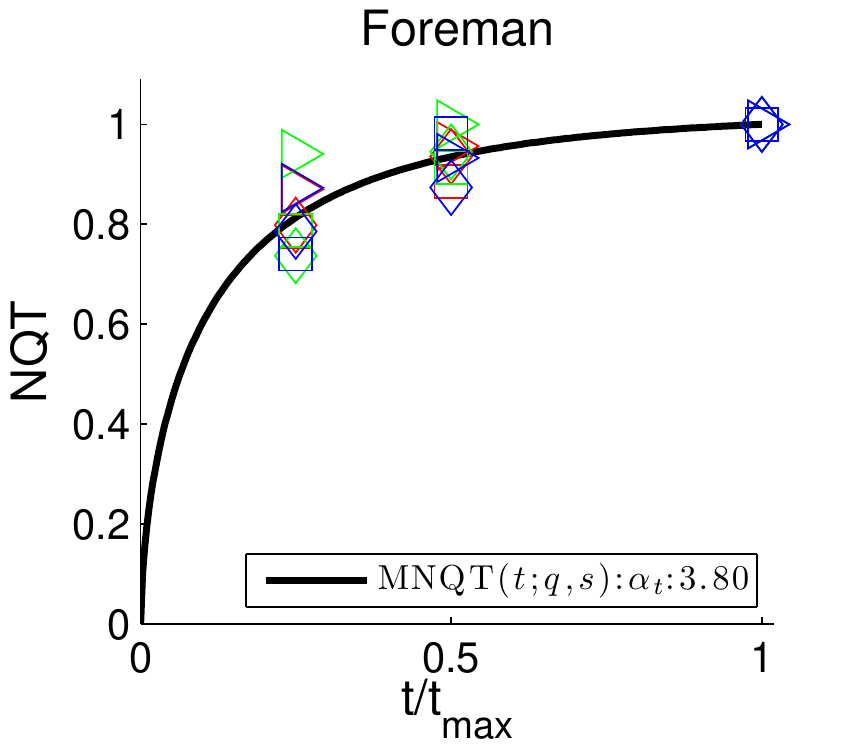}
\includegraphics[scale=0.5]{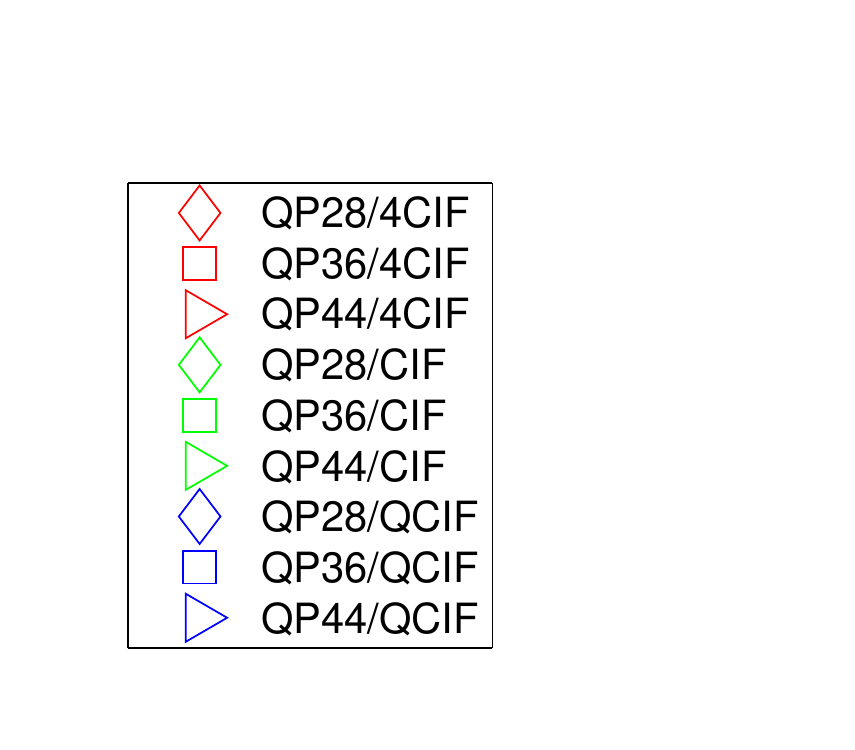}
\caption{Normalized quality v.s. normalized TR. Points are measured data under different QS's and SR's. Curves are derived from the model given in  (\ref{eq:NQT_model}) with $\beta_t$=0.63. The model parameter $\alpha_t$ is determined by least squares fitting of data at all SR's and QS's. PCC=$0.891$, RMSE=$0.052$.}
\label{fig:NMOSt}
\end{figure*}

\subsection{Modeling Normalized Quality v.s. Quantizations}\label{ssec:MNQQ}
In this subsection, we explore how QS affects the perceived quality when SR and TR are fixed. Towards this goal, we plot the normalized quality v.s. inverted normalized QS $q_{\min}/q$ (NQQ) at same SR and TR in Fig.~\ref{fig:NQQ_data}. Note that $q_{\min}/q$ can be considered as the normalized amplitude resolution. The NQQ is defined as
\begin{align}\label{eq:NQQ}
{\rm NQQ}(q;s,t) = \frac{\textrm{MOS}(s,t,q)}{\textrm{MOS}(s,t,q_{\min})},
\end{align}
where $q_{\min}$ is the minimum QS. We set $q_{\min}$ = 16, corresponding to QP = 28 in our study\footnote{Recall that even though we have tested videos coded at QP=22 at TR=30 Hz, the minimal QP that is tested at all TR and SQ is QP=28.}). In Fig.~\ref{fig:NQQ_data}, we can observe that the NQQ for different TR's but the same SR tend to cluster together. Based on both this observation and the ANOVA test on NQS in Tab.~\ref{tab:ANOVA_dependency_test}, we propose to model NQQ using a function that depends on SR, but not TR.
We further find that the generalized inverse exponential function can also characterize relation of NQQ with inverted normalized QS at the same SR, and hence propose the following model for NQQ (MNQQ)
\begin{equation}\label{eq:NQQ_model}
{\rm MNQQ}(q;s)=\frac{1- e^{-\alpha_q(s)(\frac{q_{\min}}{q})^{\beta_q}}}{1-e^{-\alpha_q(s)}},
\end{equation}
where $\alpha_q$ is the model parameter. This parameter characterizes the quality decay rate as $q$ increases, with a smaller value corresponding to a slower dropping rate. Based on the previous analysis, we assume $\alpha_q$ depends on $s$ but not $t$. We derive $\alpha_q$ for each SR for a
test sequence by least squares fitting using measured NQQ data for that SR, at all TR's. Similar to $\beta_s$ in~(\ref{eq:NQS_model}), we found that $\beta_q$=1 works well for all 7 source sequences, so that only a single parameter $\alpha_q$ is content-dependent and SR-dependent.
Figure~\ref{fig:MNQQ_model_invexp} shows that the MNQQ model is very accurate. We further evaluate the model accuracy when the parameter $\alpha_q$ is allowed to vary with TR.
Table~\ref{tab:bt_bs_dependency} shows that allowing $\alpha_q$ to vary with TR does not improve the model accuracy significantly.


\begin{table}[htp]
\centering
\caption{Model accuracy under different assumptions.}
\label{tab:bt_bs_dependency}
\small{\begin{tabular}{c|c|c|c}
\hline
Model & Assumptions &PCC&RMSE\\
\hline
\hline
\multirow{3}{*}{MNQS} & $\alpha_s$ depends on TR & 0.995 & 0.025\\
 & $\alpha_s$ independent of TR& 0.992 & 0.030\\
 & using $\hat{\alpha}_s$ in (\ref{eq:alpha_s_QP}), independent of TR & 0.989 & 0.035 \\
\hline
\multirow{2}{*}{MNQT} & $\alpha_t$ depends on SR and QS  & 0.972 & 0.026\\
 & $\alpha_t$ independent of SR and QS & 0.891 & 0.052\\
 \hline
\multirow{2}{*}{MNQQ} & $\alpha_q$ depends on TR & 0.995 & 0.020\\
 & $\alpha_q$ independent of TR & 0.982 & 0.041\\
\hline
\end{tabular}}
\end{table}
\subsection{Modeling Normalized Quality v.s. Temporal Resolution}\label{ssec:MNQT}
In this subsection, we explore how TR affects perceived quality when SR and QS are fixed. Towards this goal, we plot the normalized quality v.s. normalized TR $t/t_{\max}$ (NQT) at same SR and QS in Fig.~\ref{fig:NQT_data}. The NQT is defined as
\begin{align}\label{eq:NQT}
{\rm NQT}(t;s,q)=\frac{\textrm{MOS}(s,t,q)}{\textrm{MOS}(s,t_{\max},q)},
\end{align}
where $t_{\max}$ is the maximum TR (here, $t_{\max}$ = 30Hz). From Fig.~\ref{fig:NQT_data}, we can observe that the dropping trends of NQT for different SR's and QS's tend to cluster together and do not follow a consistent trend. Based on this observation and the ANOVA test for NQT given in Tab.~\ref{tab:ANOVA_dependency_test}, we choose to use a model function that is independent of both SR and QS.
By examining the general trend of how NQT changes with normalized TR, we propose the follow model for NQT data (MNQT)
\begin{equation}\label{eq:NQT_model}
{\rm MNQT}(t)=\frac{1- e^{-\alpha_t(\frac{t}{t_{\max}})^{\beta_t}}}{1-e^{-\alpha_t}}.
\end{equation}
The parameter $\alpha_t$ controls how fast the NQT drops as $t$ decreases, with a smaller value corresponding to a faster dropping rate. Based on the previous analysis, we assume $\alpha_t$ is independent of both SR and QS, and derive its value for each test sequence by least squares fitting using measured NQT data at all SR's and QS's. Similar to $\beta_s$ in~(\ref{eq:NQS_model}), $\beta_t$ is a constant of 0.63 for all seven sequences, which is found by least square fitting for all NQT data. Figure~\ref{fig:NMOSt} shows that the model curves can capture the data trends quite well. We also compute the PCC and RMSE of the model when using a best fitting $\alpha_t$ for each different pair of SR and QS. Table~\ref{tab:bt_bs_dependency} (middle two rows) shows that this brings slight improvement in terms of PCC and RMSE. However, considering that we already achieve high PCC and low RMSE with a parameter that is independent of both SR and QS, we choose to use this option to reduce the model complexity.


\subsection{The Overall Q-STAR Model}\label{ssec:Overall_QSTAR}
To derive the overall quality model as a function of $s$, $t$, $q$, we recognize that the normalized MOS can be decomposed  in any of the following ways:

\begin{subequations}
\begin{eqnarray}
\lefteqn{\frac{{\rm MOS}(s,t,q)}{{\rm MOS}(s_{\max}, t_{\max},q_{\min})}}\nonumber \\
= &{\rm NQS}(s; t_{\max}, q_{\min}) {\rm NQT}(t; s, q_{\min}) {\rm NQQ}(q; s, t)\\
= &{\rm NQS}(s; t_{\max}, q_{\min}) {\rm NQQ}(q; s, t_{\max}) {\rm NQT}(t; s,q)\\
= &{\rm NQT}(t; s_{\max},q_{\min}) {\rm NQS}(s;t,q_{\min}) {\rm NQQ}(q;s,t) \\
= &{\rm NQT}(t; s_{\max},q_{\min}) {\rm NQQ}(q;s_{\max},t) {\rm NQS}(s;t,q)\label{eq:QSTAR_order1} \\
= &{\rm NQQ}(q; s_{\max},t_{\max}) {\rm NQS}(s;t_{\max},q) {\rm NQT}(t;s,q)\label{eq:QSTAR_order2} \\
= &{\rm NQQ}(q; s_{\max},t_{\max}) {\rm NQT}(t;s_{\max},q) {\rm NQS}(s;t,q)\label{eq:QSTAR_order3}.
\end{eqnarray}
\end{subequations}
\begin{figure*}[htp]
\centering
\includegraphics[scale=0.5]{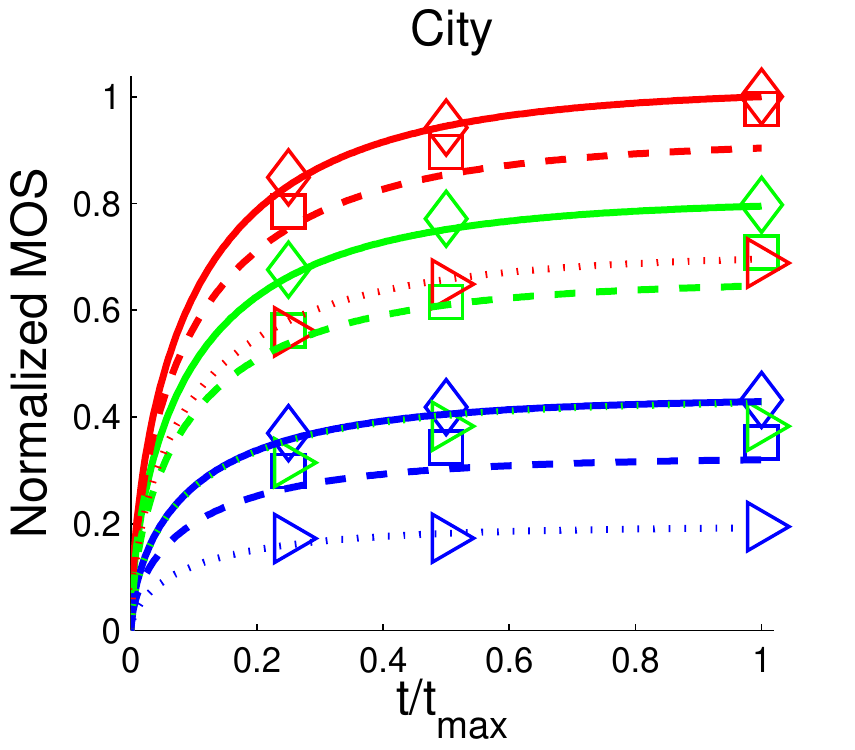}
\includegraphics[scale=0.5]{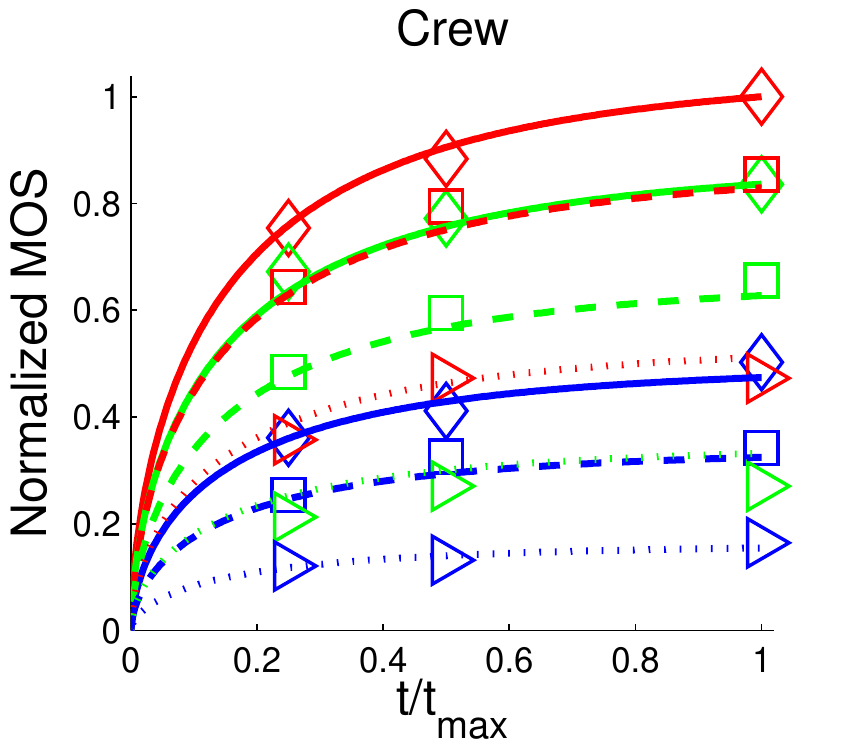}
\includegraphics[scale=0.5]{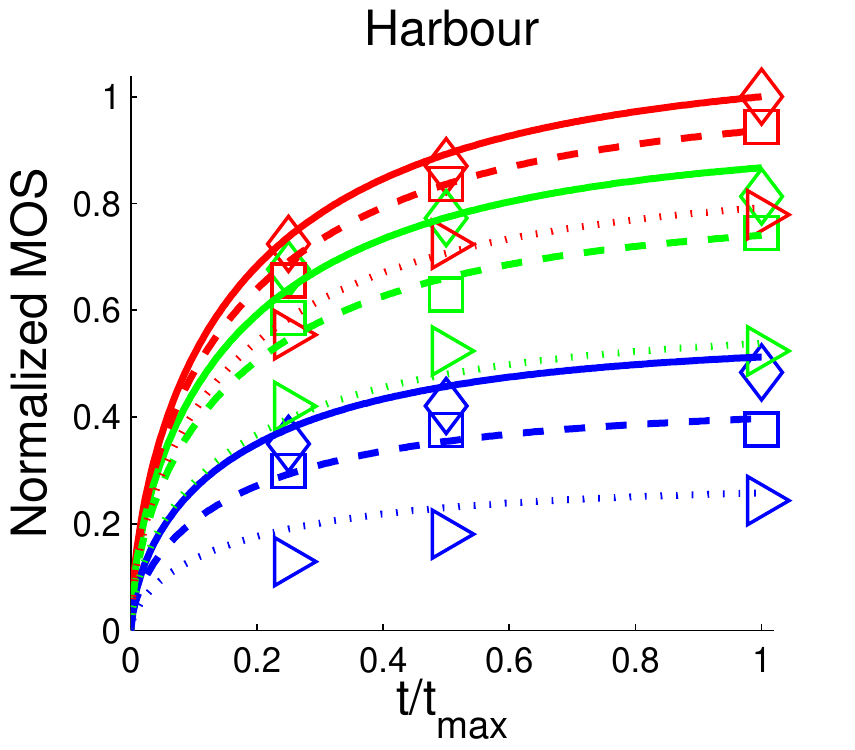}
\includegraphics[scale=0.5]{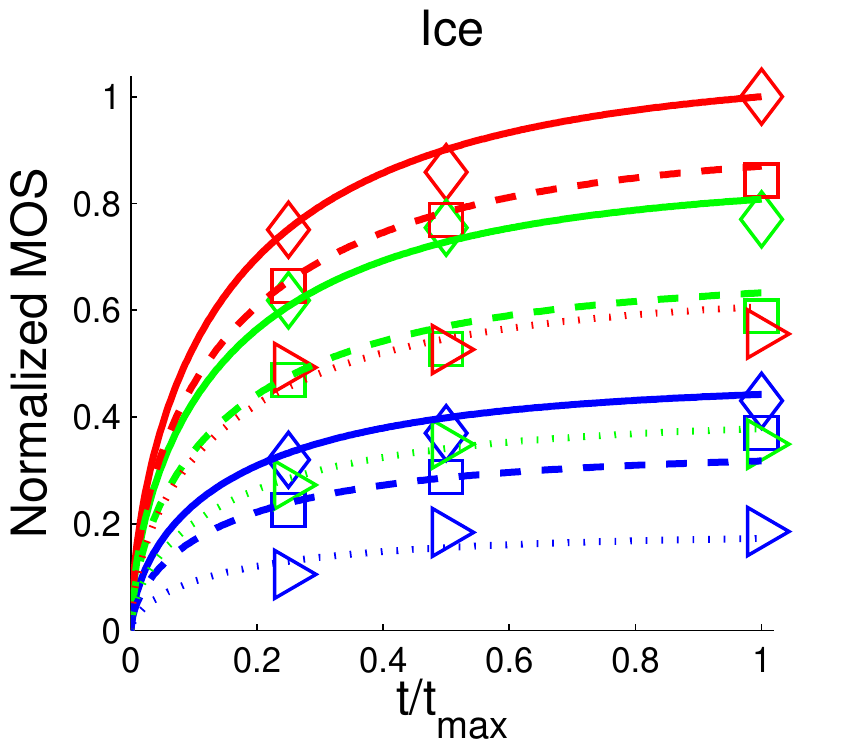}
\includegraphics[scale=0.5]{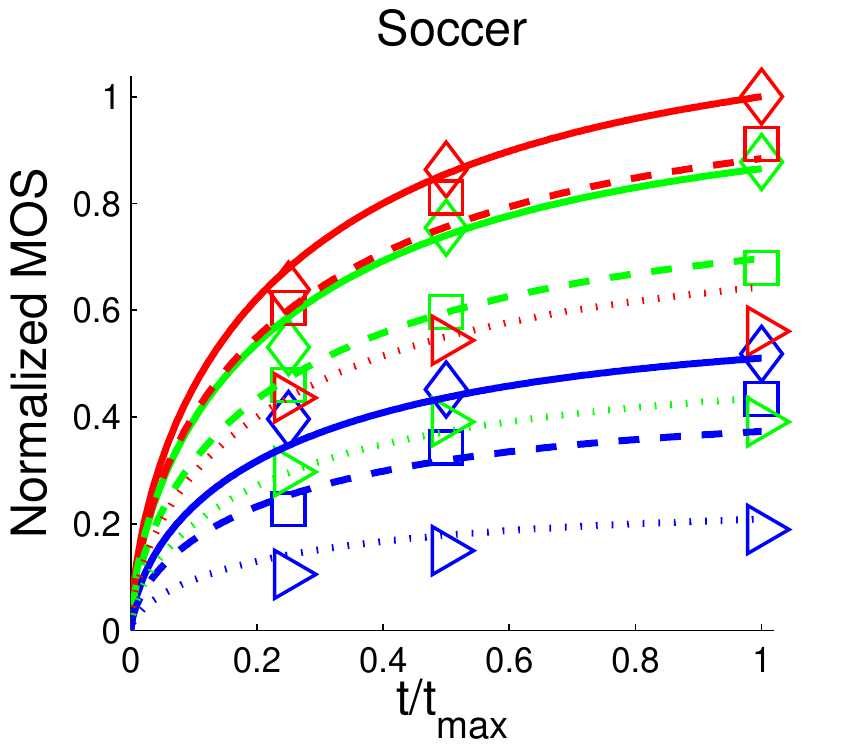}
\includegraphics[scale=0.5]{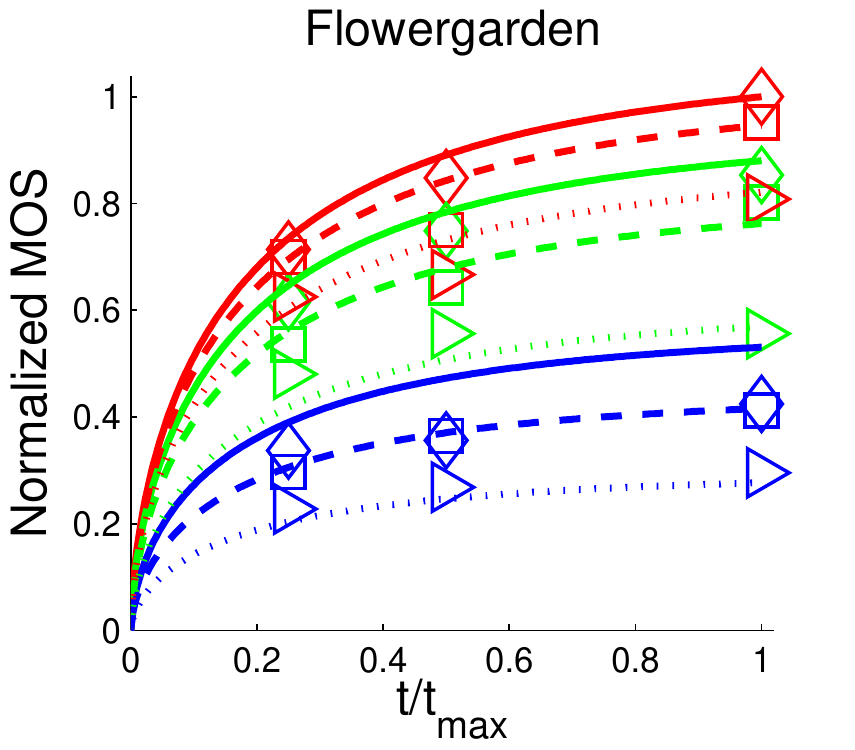}
\includegraphics[scale=0.5]{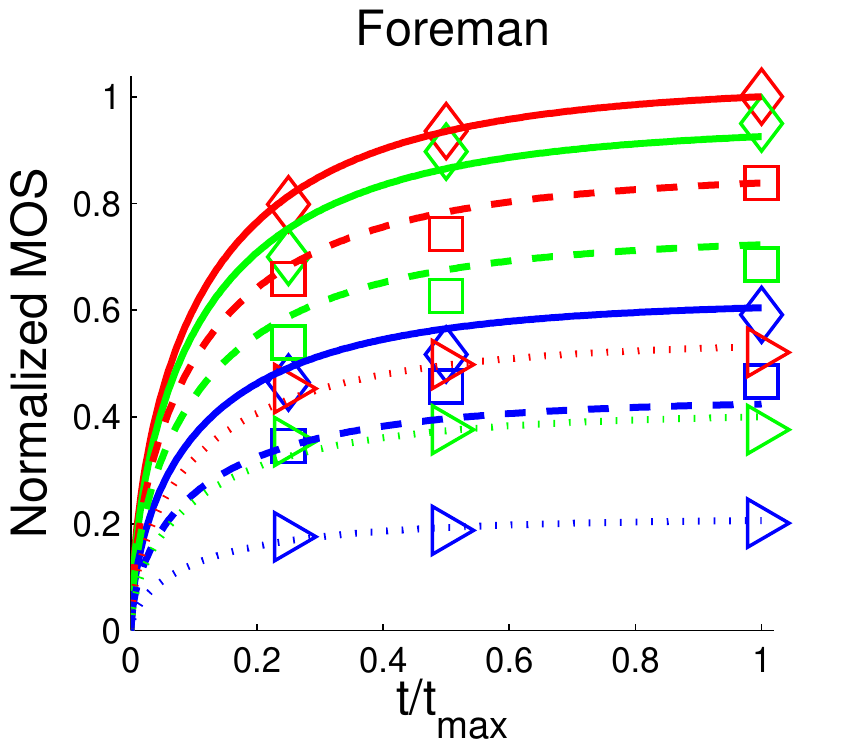}
\includegraphics[scale=0.5]{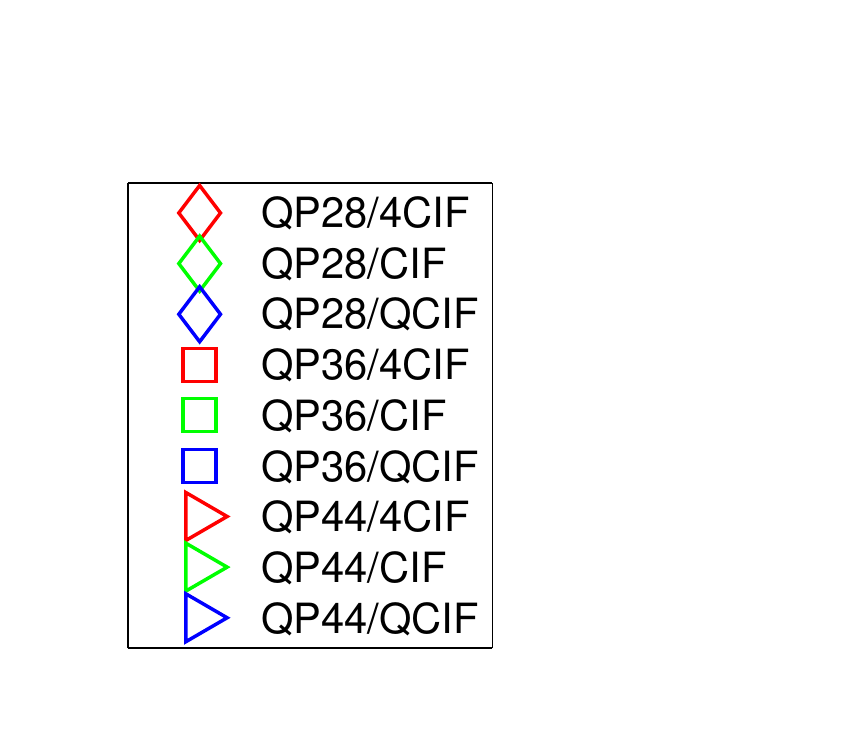}
\caption{Predicted normalized quality (in curves) and measured normalized MOS (in points) v.s. $t/t_{\max}$ under different SR's and QS's. The model parameters $\alpha_q$, $\hat{\alpha}_s$ and $\alpha_t$, are obtained by least square fitting, given in Table~\ref{tab:quality_performance}.}
\label{fig:MOSt}
\end{figure*}

\begin{figure}[htp]
\centering
 \includegraphics[scale=0.5]{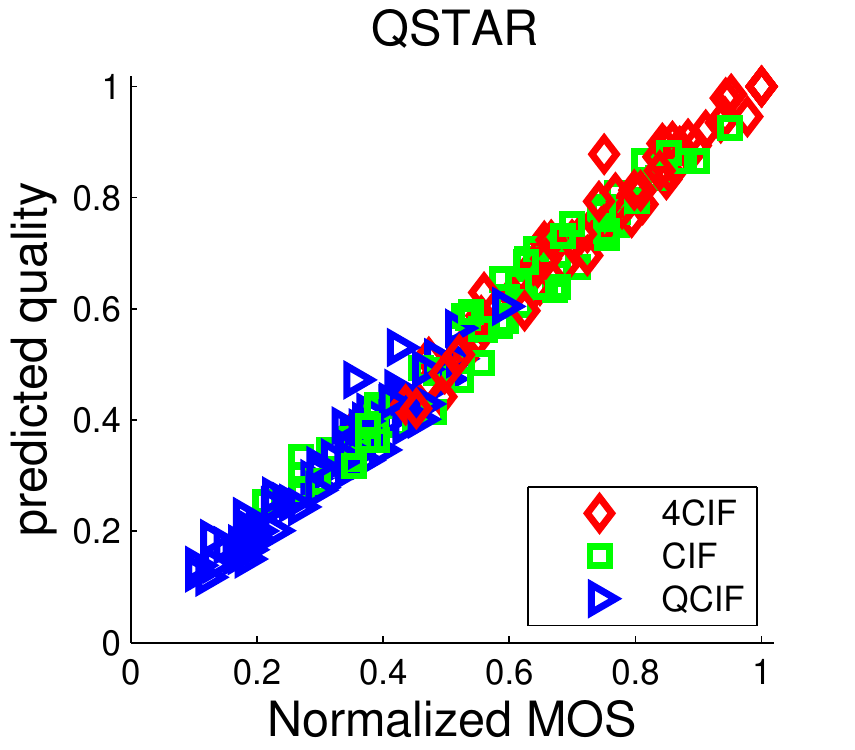}
\caption{Predicted quality using QSTAR model against measured MOS.}\label{fig:pMOS_MOS}
\end{figure}

Among these decomposition orders, we choose the one that will require the least number of model parameters while maintaining high accuracy. Because NQT term is independent of both SR and QS,  and the NQS and NQQ terms are both independent of TR, we could put NQT at any place, and it will only require a single parameter $\alpha_t$, and it will not affect the number of parameters needed for NQS and NQQ. Between NQS and NQQ,  if we choose to put  NQQ term after the NQS term, we would need to estimate $\alpha_q$ for each $s$. This is because the NQQ parameter $\alpha_q$ depends on $s$, and we don't have a good model that relates $\alpha_q$ with $s$. On the other hand, if we put the NQS term after the NQQ term, we only need to estimate $\alpha_q$ for $s$=$s_{\max}$, and because of (\ref{eq:alpha_s_QP}), we only need to estimate $\hat{\alpha}_s$ to obtain $\alpha_s$ for all $q$. Based on these considerations, we could use either (\ref{eq:QSTAR_order1}), (\ref{eq:QSTAR_order2}) or  (\ref{eq:QSTAR_order3}) to reduce the model parameters while maintain high model performance. Because NQT($t;s,q$) does not depend on $s$ and $q$, either decomposition will give the same model.
By replacing NQS, NQQ, NQT in (\ref{eq:QSTAR_order2}) with their models described in~(\ref{eq:NQS_model}), (\ref{eq:NQQ_model}) and (\ref{eq:NQT_model}), respectively, the proposed overall quality model as a function of $s$, $t$, $q$, to be called QSTAR, can be written as,
\begin{align}\label{eq:QSTAR_model}
&{\rm QSTAR}(s,t,q)={\rm MNQQ}(q; s_{\max}) {\rm MNQS}(s;q) {\rm MNQT}(t)\nonumber \\
&=\frac{1- e^{-\alpha_q(\frac{q_{\min}}{q})}}{1-e^{-\alpha_q}}\frac{1- e^{-\hat{\alpha}_sL(({\rm QP}(q))(\frac{s}{s_{\max}})^{\beta_s}}}{1-e^{-\hat{\alpha}_sL({\rm QP}(q))}}\frac{1- e^{-\alpha_t(\frac{t}{t_{\max}})^{\beta_t}}}{1-e^{-\alpha_t}},
\end{align}
where $\beta_s=0.74$, $\beta_t=0.63$ and $L(({\rm QP}(q))$ is defined in (\ref{eq:alpha_s_QP}), with $\upsilon_1$=$-$0.037, $\upsilon_2$=2.25. The model has three content-dependent parameters $\alpha_q$, $\hat{\alpha}_s$ and $\alpha_t$. We compare the predicted quality using this model with measured MOS in Fig.~\ref{fig:MOSt}, where the model parameters are obtained by least squares fitting into the data of NQQ, NQS, and NQT, respectively. As can be seen that the model matches very well with measured MOS points for most cases. Table~\ref{tab:quality_performance} (upper portion) summarizes the model parameters, RMSE and PCC of the proposed QSTAR model. In this table we also list the average 95\% confidence interval (CI) of user ratings (normalized by maximum possible rating of each source sequence) for each source sequence. We see that the RMSE of the prediction error is much lower than  the CI for all sequences. The correlation scatter plot between predicted and measured quality is presented in Fig.~\ref{fig:pMOS_MOS}.

\begin{table}[htp]
\centering
\caption{The parameters and performance of QSTAR model.}
\label{tab:quality_performance}
\hspace{-.in}
\small{\begin{tabular}{@{}c@{ }|@{~}c@{ }|@{~}c@{ }|@{~}c@{ }|@{~}c@{ }|@{~}c@{ }|@{~}c@{~}|@{~}c@{~}|@{~}c@{~}}
\hline
&city&crew&harbour&ice&soccer&fg&foreman&Avg.\\
\hline
\hline
~ &\multicolumn{8}{c}{Parameters obtained by least square fitting with MOS data}\\
\hline
$\alpha_q$& 7.25 & 4.51 & 9.65 & 5.61 & 6.31 & 10.68 & 4.57&\\
$\hat{\alpha_s}$& 3.52& 4.07& 4.58& 3.68& 4.55& 4.83& 5.94& \\
$\alpha_t$& 4.10 & 3.09 & 2.83 & 3.00 & 2.23 & 2.80 & 3.80&\\
\hline
RMSE & 0.018 & 0.025 & 0.038 & 0.033 & 0.032 & 0.058 & 0.038 & 0.035\\
\hline
$\rm PCC$ & 0.998 & 0.996 & 0.992 & 0.993 & 0.992 & 0.979 & 0.991& 0.991\\
\hline
avg. CI & 0.048 &  0.049 &  0.050 & 0.050 & 0.050 & 0.051 & 0.049 & 0.050\\
\hline
\hline
\end{tabular}}
\end{table}
\begin{table}[htp]
\centering
\caption{Three way ANOVA on MOS data.}\label{tab:ANOVA_STAR}
\begin{tabular}{|c|c|c|}
\hline
Factors & F-value & $p$-value\\
\hline
\hline
QS & 362.62 & 0 \\
\hline
SR & 698.21 & 0 \\
\hline
TR & 84.49  & 0 \\
\hline
QS $\cdot$SR & 5.5 & 0.0004 \\
\hline
QS $\cdot$TR & 3.68  & 0.0068 \\
\hline
SR $\cdot$TR & 3.77 & 0.0059 \\
\hline
QS$\cdot$TR$\cdot$SR & 0.4 & 0.921 \\
\hline
\end{tabular}
\end{table}

\begin{table}[htp]
\centering
\caption{Three way ANOVA test on NQS, NQT, and NQQ data, respectively.}\label{tab:ANOVA_dependency_test}
\begin{tabular}{|c|c|c|c|}
\hline
~ & Factors & F-value & $p$-value\\
\hline
\hline
\multirow{4}{*} {NQS} & QS & 1940 & 0\\
& TR & 0.24 & 0.78 \\
& SR$\cdot$TR & 0.68 & 0.64 \\
& SR$\cdot$QS & 20.68  & 0 \\
\hline
\multirow{4}{*}{NQT}& QS &3.82 & 0.02\\
& SR & 5.59 & 0 \\
& TR$\cdot$SR & 1.49 &  0.21 \\
& TR$\cdot$QS & 2.21 & 0.30 \\
\hline
\multirow{4}{*}{NQQ} & SR & 24.08 & 0\\
& TR & 0.22 & 0.8 \\
& QS$\cdot$SR & 8.77 & 0 \\
& QS$\cdot$TR & 0.37 & 0.83 \\
\hline
\end{tabular}
\end{table}

\begin{table}[htp]
\centering
\caption{Data Set Description} \label{tab:datasets}
\begin{tabular}{|p{2 cm}|p{6 cm}|}
 \hline
 {\tt \small{DataSet\#1}} (this paper) & 7 source sequences in 4CIF/30Hz coded at 3 frame rates (30, 15, 7.5 Hz), 3 spatial resolutions (4CIF, CIF, QCIF), and 3 QP levels (28, 36, 44). A total of 189 PVSs. Videos are coded using H.264/SVC and displayed on a mobile device in full screen.\\
\hline
{\tt \small{DataSet\#2}}~\cite{demin} & 5 source sequences in CIF/30Hz coded at 3 frame rates (30, 15, 7.5 Hz), five different bit rate levels (each frame rate has its corresponding 5 QPs). A total of 75 PVSs. Videos are coded using H.263 and displayed on a laptop device at the native spatial resolution.\\
 \hline
 {\tt \small{DataSet\#3}}~\cite{cheon} & 3 source sequences used in CIF/30Hz coded at 3 frame rates (30, 15, 7.5 Hz), 6 spatial resolutions (in between QCIF and CIF), and 3 bit rate levels. A total of 54 PVSs. Videos are coded using H.264/SVC and displayed on a desktop device at native spatial resolution.\\
 \hline
{\tt \small{DataSet\#4}}~\cite{HDTV_vqeg} & 9 1080i HDTV source sequences coded at 3 and 4 QP levels (or bit rates) by H.264 and MPEG-2, respectively. Only test sequences without packet loss are presented. A total of 63 PVSs. Videos are displayed on a HDTV monitor at the native spatial resolution.\\
 \hline
 {\tt \small{DataSet\#5}}~\cite{PairedComp_JS} & 3 720p/50Hz source sequences coded at 3 spatial resolutions (1280$\times$720, 640$\times$360, 320$\times$180), 4 frame rates (50, 25, 12.5, 6.25) and several bit rate levels (controlled by varying QS) by H.264/SVC. A total of 26 PVSs. Videos are displayed on a LCD monitor at 1280$\times$720 spatial resolution.\\
 \hline
 {\tt \small{DataSet\#6}}~\cite{yenfu_csvt} & 7 CIF/30Hz sequences coded at 4 frame rates (30, 15, 7.5, 3.75 Hz) and 3 QP levels (28, 36, 40). A total of 100 PVSs. Videos are coded using H.264/SVC with joint temporal and amplitude scalability and displayed on the laptop device in laptop screens in native resolution with gray-out background.\\
 \hline
\end{tabular}
\end{table}



\section{Statistical Analysis of Subjective Test Results}\label{sec:ANOVA_analysis}
To examine the statistical significance of the effect of SR, TR, and QS and their interactions on the MOS data, we conduct three way ANOVA test, which computes the probability ($p$-value, which is derived from the cumulative distribution function of F based on the F-value) of the null hypothesis that the differences in the MOS values due to the changes in the examined variables is due to chance. If this probability is low (i.e. $p$-value $<$ 0.05), we consider the examined variable as having statistically significant influence on MOS. Otherwise, we say that the examined variable has statistically insignificant influence on MOS.

As shown in Tab.~\ref{tab:ANOVA_STAR}, SR, TR, and QS each has significant impact on the MOS.  Furthermore, there are statistically significant interaction between SR and TR, between SR and QS, and between TR and QS on the quality ratings.  However, there is no significant interaction among all three variables.
The above test examines the impact of the STAR on the absolute quality ratings directly. To further examine whether SR and TR influence the quality drops as QS increases, when SR and TR are fixed, we perform three-way ANOVA test on the NQQ data as well as the interaction of QS with SR and TR. Similarly we perform three-way ANOVA tests on NQS and NQT data.  These results are reported in Tab.~\ref{tab:ANOVA_dependency_test}.  The ANOVA test for NQS reveals that QS has significant effect on NQS, but TR doesn't; furthermore, the interaction between SR and QS is significant, but not the interaction between SR and TR. Note that the insignificant interaction between SR and TR essentially tells us  that the dropping rate of NQS curve as SR decreases does not depend on TR in a statistically significant way.
Likewise, the ANOVA test for NQQ reveals that SR has significant effect on NQQ, but TR doesn't; furthermore, the interaction between QS and SR is significant, but not the interaction between QS and TR. Different from the results for NQS and NQQ, the ANOVA test for NQT reveals that both SR and QS have significant impact on NQT, but the interaction between TR and SR, and that between TR and QS are both insignificant. This suggests that the differences in the dropping rates of NQT curves under different SR and QS are mostly caused by viewer rating noise. These results are consistent with the observations made earlier about the NQS, NQT, and NQQ data, and support the assumptions we made while deriving the QSTAR model (i.e. $\alpha_s$ depends on QS but not TR, $\alpha_q$ depends on SR but not TR, and $\alpha_t$ is independent of SR and QS).
These results are consistent with the observations made earlier about the NQS, NQT, and NQQ data, and support the assumptions we made while deriving the QSTAR model (i.e. $\alpha_s$ depends on QS but not TR, $\alpha_q$ depends on SR but not TR, and $\alpha_t$ is independent of SR and QS).

Note that with the default deblocking filter in H.264, higher QS causes more noticeable blurring effect than typical quantization effect. Because we display PVS's at the full screen resolution (with spatial interpolation when necessary), lower SR sequences also exhibit significant blurring.  We suspect that the interaction of SR and QS on both NQQ and NQS is  because  reducing SR and increasing QS both can  lead to increased blurring effect, making it harder to separate the perceived distortion due to higher QS and those due to lower SR. On the other hand, reducing TR causes unsmooth rendering of fast moving objects, which is an artifact that is quite distinguishable from the blurring artifacts associated with lower SR or higher QS. This may be the reason that  there is no significant interaction between SR and TR on both NQS and NQT, and between QS and TR on both NQT and NQQ.

\section{Validation of Proposed Models}\label{sec:Validation_model}
\subsection{Model Validation over Other Datasets}
In order to verify the accuracy of the model form of QSTAR model on other datasets, we apply the model partially or fully on five other datasets reported in~\cite{demin, cheon,HDTV_vqeg,PairedComp_JS,yenfu_csvt}. Brief description of these datasets are given in Tab.~\ref{tab:datasets}. {\tt \small{DataSet\#1}} refers to the one used in this paper to train the proposed model.

Note that only {\tt \small{DataSet\#3}} and {\tt \small{DataSet\#5}} containing PVS's varying in all three resolution dimensions,  but the ranges of the SR different from the range we examined ({\tt \small{DataSet\#1}}). For {\tt \small{DataSet\#2}} and 6, which contain PVS's in different TR and QS, but all at CIF resolution, we validate a special case of QSTAR: QSTAR($t, q$; CIF) = MNQQ($q$; CIF) MNQT($t$). In addition to trying to fit the overall model QSTAR($s,t,q$) or QSTAR($t,q$; CIF) to the entire set of MOS data at different combinations of ($s,t,q$) for a database, we also try to validate the individual model (i.e., MNQS, MNQT, MNQQ) when there are sufficient testing points. For instances, when there are multiple PVS's at different TR's under the same SR and QS, we can validate the MNQT model. {\tt \small{DataSet\#4}} contains sequences at the same SR (HDTV) and TR (30Hz) but different QS's. We only validate the MNQQ model in this case, since only this dataset provides the QP value.
While applying the proposed model, we use the same values for the shaping parameters $\beta_s$, $\beta_t$, $\beta_q$ that are derived from our training data {\tt \small{DataSet\#1}} (as reported in Sec.~\ref{sec:QSTAR}), but apply the optimal values for $\hat{\alpha}_s$, $\alpha_t$, $\alpha_q$ for each source sequence, derived by least squares fitting with the given MOS data for that sequence at available ($s,t,q$) combinations.

Table~\ref{tab:all_model_validation} summarizes the validation performance in terms of PCC and RMSE. It can be seen that QSTAR model (or its reduced version under the same SR) has high PCC ($>$0.85) and low RMSE for all databases examined.
In addition, both MNQQ and MNQT models have  high correlation with all applicable datasets.
Furthermore, we have observed in several datasets, NQQ dropping rate is independent of TR ({\tt \small{DataSet\#2}}, {\tt \small{\#6}}), and NQT dropping rate  is independent of QS ({\tt \small{DataSet\#6}}). These trends are consistent with our observations from {\tt \small{DataSet\#1}}.



\subsection{Discussion About the Applicability of The Model}
\subsubsection{Choice of Function Forms}
Besides the inverted exponential function, we also investigated using other functional forms, such as power law, logarithm functions,
to model NQQ and NQT. We have found that the inverted exponential function with a properly chosen $\beta$ can model our dataset as well as others more accurately.  In our earlier work based on {\tt \small{DataSet\#6}}~\cite{yenfu_csvt}, we proposed to model NQT using the  inverse exponential function with $\beta$=1.0. We have found that using $\beta$=0.63 (as determined from {\tt \small{DataSet\#1}}) in fact fits NQT data in {\tt \small{DataSet\#6}} slightly better (PCC=0.98 vs. 0.95). In~\cite{yao2}, also based on {\tt \small{DataSet\#6}}, we proposed to model NQQ using an exponential decay function of $q/q_{\min}$.  Although using the NQQ model proposed here (an inverse exponential function of $q_{\min}/q$ with $\beta$=1.0)  yields very similar performance (PCC=0.995 vs. 0.991),  we believe the NQQ model proposed here is generally more appropriate because it correctly characterizes the initial slow drop when the QS is only slightly larger than $q_{\min}$.

\subsubsection{Alternative models for NQQ}
In this paper, the NQQ is modeled in terms of QS. Noting that QS affects the PSNR of decoded frames, in our earlier work using {\tt \small{DataSet\#6}}~\cite{yenfu_csvt}, we have used the average PSNR of decoded frames to model the NQQ data, and we have found that the NQQ can be modeled accurately using a sigmoidal function of PSNR.  We have found that the same model is applicable to the present dataset as well.  We further find that NQQ at a given SR and TR can be modeled accurately using an inverse exponential function of the bit rate, normalized by the bit rate required when using $q_{\min}$ under the same SR and TR.  Each NQQ model has its unique advantages. The QSTAR model using the NQQ v.s. QS model, together with a rate model that relates the bit rate with STAR, enables the encoder or a scalable video adaptor to optimize the  QS, SR and TR for a given target rate. On the other hand, the QSTAR model using PSNR or bit rate to characterize  the NQQ function allows quality evaluation at the decoder (especially when the QS is not available or the QS is time varying). We further note that the NQQ model in itself is useful when a video is coded at a fixed spatial and temporal resolution.

\subsubsection{Normalized Quality}
The proposed model characterizes the normalized quality, relative to the maximum quality achievable at a highest STAR.  In practical applications,  the maximum affordable STAR (and hence the maximum possible quality) at a particular video receiver is often fixed due to its bandwidth and display capacity.  It is the relative quality compared to this maximum that is of importance. The absolute quality may be of importance
only when the system designer has to allocate a common resource (e.g. total bandwidth) over multiple videos, and when these videos could have quite different absolute qualities even when they are all coded at the highest STAR.

\subsubsection{Range of STAR}
The model proposed here is developed based on the range of STAR considered in our subjective tests, i.e., QP from 28 to 44, SR from QCIF to 4CIF (similar to VGA), and TR from 7.5 to 30Hz. Under these ranges, we have found that NQQ, NQS, and NQT can each be modeled well using the inverse exponential function with a properly chosen shaping parameter. This function (when $\beta<$ 0.63) predicts the quality will drop increasingly faster as the QS increases and SR and TR decrease.  On the other hand, it is likely that the quality will saturate to a very low level once the STAR goes below a certain level. In other words, the model may not be valid in the range outside the low saturation point.  However, we think that the QSTAR model is appropriate for the range of STAR likely to be used for practical applications. Validation with other datasets presented in this section  confirms that the model is accurate  for  even larger SR  and TR ranges .  Our preliminary subjective tests (not included in {\tt \small{DataSet\#1}}) have indicated that using QP lower than 28 does not provide easily noticeable quality improvement, whereas using QP higher than 44  leads to unacceptable video quality.

\subsubsection{Influence of Display and Encoder Settings}

The datasets considered in our validation study are obtained under quite different display environments and using different encoders. The fact that the proposed model is quite accurate for these datasets suggest that the proposed model function is generally applicable, although the model parameters depend on the display environment (including the spatial and temporal interpolation filter used for display lower SR/TR video in full resolution) and encoding algorithm. For adoption of these models in practical applications, one may need to conduct subjective tests for  the intended display environment and encoder setting, and derive appropriate model parameters in advance.

\begin{table}[!ht]
\centering
\caption{Performance of QSTAR model of different databases.}
\label{tab:all_model_validation}
\begin{tabular}{@{~}c@{ }|@{~}c@{ }|@{~}c@{ }|@{~}c@{ }|@{~}c@{ }|@{~}c@{ }|@{~}c@{ }|@{~}c@{ }}
\hline
\hline
Model & Metrics & {\tt \footnotesize{DataSet}\#1}&{\tt \small{\#2}} &{\tt \small{\#3}} &{\tt \small{\#4}} &{\tt \small{\#5}}&{\tt \small{\#6}}\\
\hline
\multirow{2}{*}{QSTAR($s,t,q$)}& RMSE & 0.035 & - & 0.039 &- & 0.080 &- \\
\cline{2-8}
& PCC & 0.991 & - & 0.959 & - & 0.856 &- \\
\hline
\multirow{2}{*}{QSTAR($t, q$;CIF)}& RMSE & -&0.120 & 0.023 & - & - & 0.044 \\
\cline{2-8}
& PCC & -& 0.854 & 0.975 & - & - & 0.973 \\
\hline
\multirow{2}{*}{MNQQ($q$)}& RMSE & 0.041 & 0.052 & 0.022 &0.029 & 0.079 & 0.046 \\
\cline{2-8}
& PCC & 0.982 & 0.961 & 0.980 & 0.993 & 0.819  & 0.963 \\
\hline
\multirow{2}{*}{MNQT($t$)}& RMSE & 0.052 & 0.030 & 0.066 &- & 0.032 &0.034  \\
\cline{2-8}
& PCC & 0.891 & 0.964 & 0.939 & - &0.987  &0.968 \\
\hline
\multirow{2}{*}{MNQS($s$)}& RMSE & 0.030 & - & 0.049 & - & - & -\\
\cline{2-8}
& PCC & 0.992& - & 0.947 & - & - & - \\
\hline
\hline
\end{tabular}
\end{table}

\section{Conclusion}\label{sec:conclusion}
In this work, we propose a perceptual quality model considering the impact of SR, TR and QS  based on subjective tests conducted on a mobile display platform. In this model, we use a one-parameter function to capture the quality decay v.s. SR, TR and QS individually. The parameter in each function is sequence dependent.
Interestingly, we found that the dropping rate of the quality with TR, characterized by $\alpha_t$, is independent of SR and QS, and the dropping rate of quality with both SR and QS, indicated by $\alpha_s$ and $\alpha_q$, respectively, are both independent of TR. Although the dropping rate with SR  $\alpha_s$ is dependent on QP, we found that they are related linearly. The overall model only requires  three content-dependent parameters.
We further applied the proposed model over five other datasets, which contain subjective ratings for different source videos compressed using different encoders and displayed under different environments. Both the overall QSTAR model and the individual models MNQT and MNQQ are shown to be quite accurate (PCC$>$0.8 in all cases). Based on these results, we conclude that the proposed QSTAR model and its components MNQQ, MNQT, and MNQS are applicable to various encoding and display environments, but the model parameters generally depend on the encoding and display setting.

In this work, we only emphasize derivation of a simple but accurate model form of QSTAR in the absent of content-derived model parameters. We believe that the model parameters are correlated with the video content very well, so that a comprehensive method to automatically predict the model parameters from content features is essential. In the on-going research, we will investigate how to estimate parameters from underlying source sequences so that the QSTAR model can be more complete.

In another subjective test~\cite{yuan_icip2012}, we have compared scalable video coded using H.264/SVC and non-scalable video produced using H.264/AVC, under the same STAR, using pair comparison. The study shows that their perceptual quality are very similar. This is very promising, indicating that the proposed QSTAR model is applicable to both scalable and non-scalable video.

It is worth noting the implication of the proposed model form in (\ref{eq:QSTAR_model}). It suggests  that the  quality of a video is the product of a spatial quality factor (jointly determined by SR and QS) and a temporal quality factor (determined by TR). The spatial term is in turn the product of  two factors, MNQQ and MNQS. MNQQ describes how QS affects the quality when the video is at the maximum SR; and MNQS accounts for the  quality degradation due to lower SR.  Finally the temporal quality factor or MNQT accounts for the quality degradation due to lower TR.

The proposed quality models, together with the rate model, also as a function of STAR in~\cite{zhan_rate_technical_rep}, can be used to determine the optimal STAR that maximizes the quality given a rate constraint, both for video encoding/transcoding and for scalable video adaptation. Our prior work ~\cite{yao2, RQpar_icip11_zhan} has investigated a subset of this problem, where SR is fixed, and only TR and QS are adapted, based on quality and rate models as functions of TR and QS only.  Extension of this work to includes the SR dimension, using the newly developed quality and rate models, both as functions of SR, TR, and QS,  is another interesting direction for future research, and preliminary results are presented in~\cite{Hao_icip2012}.
\section{Acknoledgement}
The authors would like to thank Dr. Zhan~Ma for his valuable contributions to the earlier part of this work and Meng Xu for his helping with feature extractions in this paper.


\end{document}